Drug delivery in tumors is enhanced by bacterial proteolytic activity in a size dependent manner –A mechanistic understanding of combination therapy

Hiroaki Shirai[a]*, Kosuke Tsukada[a]

[a]Graduates School of Science and Technology, Keio University, 3-14-1 Hiyoshi Kohoku-ku, Yokohama-shi, Kanagawa, Japan 223-8522

*Corresponding author

Hiroaki Shirai

Tel: +81 45 566 1513

Fax: +81 45 566 1587

Email: hs7796@keio.jp

Address: Keio University, 3-14-1 Hiyoshi Kohoku-ku, Yokohama-shi, Kanagawa, Japan 223-8522

Abstract

The use of bacteria has been attractive to cancer researchers as drug delivery vehicle because motile bacteria are able to penetrate in tumors. In particular, the combination of therapeutic bacteria and conventional chemotherapy leads to dramatically high anti-tumor efficay. However, the mechanisms of the synergy, in part, remain unclear. To aim for understanding the mechanisms of the synergy of the combination therapy, simultaneous delivery of *C. novyi-NT* and chemotherapeutic agents in tumors is mathematically modeled from porous media approach. Simulated doxorubicin concentration in tumors after Doxil administration with or without bacteria agreed reasonably well with experimental literature. The simulated doxorubicin concentration in tumors by the combination of Doxil and *C. novyi-NT* is over twice higher than that of Doxil alone, as observed in previous experimental literature. This enhanced concentration is because of the degradation of extracellular matrix of collagen by bacterial proteolytic activity, which reduced interstitial fluid pressure in tumors by increasing hydraulic conductivity of interstitium, and thus increases convection through vessel walls. Additionally, solid stress alleviation caused by collagen degradation increases vessel density by decompressing blood vessels. On the other hand, the simulated doxorubicin concentration in tumors for non-liposomal doxorubicin is not enhanced by *C. novyi-NT* because vascular permeability of free-doxorubicin is larger than Doxil, and thus, increased but relatively small convection across vessel walls is outweighed by the efflux due to increased interstitial flow. A strategy to further enhance this combination therapy is discussed with sensitivity analysis.

## 1. Introduction

With the increase in cancer incidence worldwide (1), developing more effective cancer treatment strategy is urgent. Chemotherapeutic agents such as anti-cancer drugs and drug-constraining liposomes do not penetrate in tumors effectively, and this is one of the major limitations of chemotherapy (2). Poor drug penetration in tumors is mainly caused by two factors: (i) high interstitial fluid pressure and (ii) dense extracellular matrix (ECM) of collagen. Interstitial fluid pressure is high in tumors because of leaky blood vessels and abnormal lymphatics in tumors (3). Leaky blood vessels allow larger molecules to infiltrate from vessels to tumor tissues, and thus, leads to high osmotic pressure in tumors. Moreover, lymphatic vessels are absent in tumors (4), which increases hydrostatic pressure of tumors. Consequently, drug delivery in tumors relies on not convection from blood vessels to tissues but diffusion (3). Moreover, penetration of larger particles such as antibody or liposomes is inhibited by dense extracellular matrix (ECM) of collagen because collagen content of extracellular matrix of tumors is high and diffusivity of larger molecules such as antibody or liposomes is negatively correlated to collagen content of tumors (5).

To overcome this limitation of conventional chemotherapy, the use of anaerobic bacteria has been attractive to cancer researchers as motile bacteria have a propelling force using their flagellar, and thus, are able to penetrate in avascular tumor necrotic regions (6) (7). Additionally, anaerobic bacteria such as *Salmonella* or *Clostridium* grow and survive in only oxygen-depleted hypoxic areas (8) (9). For example, *Bifidobacterium longum* selectively localized to and proliferated in rat mammary tumors after systemic application (10). Thus, tumor hypoxic regions provide a preferable niche for bacterial growth. These opportunities allowed the emergence of a field so-called *Bacterial Cancer Therapy,* which ranges pro-drug (9) (11) (12), drug delivery vehicle (13) (14), immunotherapy (15) (16), combination therapy (14) (17) (18) (19) (20). For example, *Clostridium novyi-NT* (*C. novyi-NT*), which is a *C. novyi* strain devoid of its lethal toxin, has been investigated for its potential because of its ability to lysis cancer cells. Intravenously injected *C. novyi-NT* spores germinated within the avascular regions of tumors in mice and destroyed surrounding viable tumor cells (19) (21). Moreover, magneto-aerotactic bacteria MC-1 cells bearing covalently bound drug-containing nanoliposomes were injected near the tumor in severe combined immunodeficient beige mice and magnetically guided, up to 55% of MC-1 cells penetrated into hypoxic regions of HCT116 colorectal xenografts (13). Furthermore, attenuated

*Salmonella* strains were genetically constructed to synthesize a therapeutic agent intracellularly and periodically lyse to release the compound into tumors (14).

Notably, therapeutic bacteria alone not only demonstrate high anti-tumor effect but the combination with conventional chemotherapy leads to dramatically high efficacy (14) (17) (18) (20) (23) (22) (24) (25) (26) (27) (28). Table 1 summarizes the previous experimental works of combination therapy of conventional chemotherapy and bacterial cancer therapy. For example, administration of *C. novyi-NT* spores together with conventional chemotherapeutic drugs lead to extensive hemorrhagic necrosis of tumors often developed within 24 h, resulting in significant and prolonged anti-tumor effects (17). This mechanism is considered to be because anti-cancer agents act on the cancer cells close to tumors while *C. novyi-NT* destroy cancer cells in tumor necrotic regions (18). The combination of *Salmonella choleraesuis* and anti-cancer agent, cisplatin, acted additively to retard tumor growth and extensively prolong the survival time of the mice bearing hepatomas or lung tumors (22). Additionally, the combination of both circuit-engineered *Salmonella* and 5-fluorouracil (5-FU) lead to a notable reduction of tumor activity along with a marked survival benefit over either therapy alone (14). Moreover, the combination therapy of *S. typhimurium VNP20009* and endostatin, an angiogenesis inhibitor, enhanced anti-tumor effects by inducing greater growth inhibition (23). Furthermore, the combination of Doxil, a *PEG*ylated liposomal doxorubicin, and *C. novyi-NT* spores resulted in complete regression of tumors in 100% of mice and 65% of the mice were still alive at 90 days, though neither doxorubicin nor Doxil resulted in prolonged therapeutic effects in the mice. Perhaps more importantly, doxorubicin concentration in tumors is four to five times enhanced by *C. novyi-NT* compared with Doxil alone (20). This enhanced concentration seemed to be due to liposome-disrupting ability of *C. novyi-*

*NT* because the doxorubicin concentration was not enhanced by the combination of free-doxorubicin and *C. novyi-NT* (20). In their work, however, for the mechanisms of this enhanced doxorubicin concentration in tumors, the possibility of enhanced extravasation from blood vessels was excluded because doxorubicin concentration after non-liposomal doxorubicin treatment was not enhanced (20). But there are several factors influencing drug extravasation from vessels and dependent on each drug type such as vascular permeability (drug diameter), binding affinity, and pharmacokinetics. Thus, the possibility of enhanced extravasation of drugs from blood vessels by the combination of chemotherapy and *C. novyi-NT* might be overlooked. Therefore, we aim for a mechanistic understanding of enhanced drug concentration by the combination of chemotherapy and bacterial cancer therapy. The objective of this work is to mathematically model simultaneous transport of chemotherapeutic agents and *Clostridium* in tumors.

Mathematical modeling of delivery of *Clostridium* in tumors is missing in literature. Mathematical modeling of drug delivery in tumors has been established and well-reviewed by Jain and co-workers (29) (30) (31) (32). Mok and coworker developed a mathematical model to describe the spread of herpes simplex virus from the initial injection site (34). Little has been discussed about the role of bacterial proteolytic activity on extracellular matrix so far. Behave and co-workers found that the spheroid morphology was lost and the cells were loosely packed by heat-inactivated *Clostridium sporogenes* and they inhibit the proliferation of cells by the morphological changes caused by them in the spheroid. This suggested that heat-inactivated bacteria inhibit the proliferation of cells by breaking down the extracellular matrix (ECM) and destroying the cell-ECM interactions. It was also suggested that this could be due to the extracellular proteases of *C.*

*sporogenes*, one of which is collagenase, that degrade the tumor tissue (35).

This work hypothesizes that bacterial proteolytic activity of *C. novyi-NT* degrades extracellular matrix of collagen in tumors, which increases hydraulic conductivity of interstitium and thus, reduces interstitial fluid pressure. Most of anaerobic bacteria such as *C. novyi-NT and Salmonella typhimurium* secrete collagenase that degrades type I, II, and III collagen (table 2). Additionally, hydraulic conductivity of interstitium negatively correlated with collagen or glycosaminoglycan (GAG) content (36) (37) (figure S1). Reduced interstitial fluid pressure in turn increases convection through vessel walls from blood vessel to tumor tissues, which thus increases drug concentration in tumors. Previous experimental work showed that collagenase treatment reduced interstitial fluid pressure and it increased the trans-capillary pressure gradient, inducing a 2-fold increase in the tumor uptake and improving the distribution of the monoclonal antibody (38). Additionally, solid stress, a physical force in the tumors by dense extracellular matrix of both fibril collagen and swelling hyaluronan, has been known to be involved in cancer therapeutic efficacy by compressing blood vessel in tumors (39). Decompressing blood vessels by depleting collagen and/or hyaluronan can improve drug delivery because it improves blood vessel perfusion or vessel density (39). Therefore, degradation of collagen by bacterial proteolytic activity alleviates solid stress of tumors, which increases vessel density and thus, increases transport of drugs across vasculature.

Thus, this work specifically hypothesizes that bacterial proteolytic activity influences transport of chemotherapeutic agents in tumors by degrading extracellular matrix (ECM) of collagen via two-pathways: (i) degradation of ECM of collagen increases hydraulic conductivity of interstitium, and thus, reduces interstitial fluid pressure, (ii) it alleviates solid stress of tumors, which increases vessel density. From these two pathways, the convective transport across tumor vasculature is enhanced, and thus, drug delivery is enhanced. This work also hypothesizes that this enhanced drug delivery in tumors is dependent on size of drugs because vascular permeability of liposomal doxorubicin with diameter 85 nm is six times smaller than free-doxorubicin (41). To validate the hypotheses above, we mathematically model the simultaneous delivery of *Clostridium* and chemotherapeutic agents in tumors. The objective of this work is to mathematically model transports of interstitial fluid, chemotherapeutic agents, and *Clostridium* in tumors. Our goal is to understand the mechanisms of the synergetic effect of the combination of chemotherapeutic agents and therapeutic bacteria; the specific goals are two-fold: (i) to understand how bacterial proteolytic activity interacts with tumor microenvironment of extracellular matrix of collagen, interstitium hydraulic conductivity, interstitial fluid pressure, and solid stress, and (ii) how remodeled tumor microenvironment influences diffusive and convective transport across vascular walls and interstitial fluid flow, and thus, enhances drug delivery.

3. Results and discussions

Interstitial fluid pressure is high at the core of tumors and drops at the periphery of the tumors (figure 2**a)**. These simulation results agree with experimental literature showing the high interstitial fluid pressure in *s.c.* adenocarcinoma at the center of tumors and low at the periphery of tumors (50). Interstitial fluid pressure in healthy liver is approximately –2.2 mmHg (51), which is roughly close to the simulated value in normal tissues at 2 mmHg (Fig. 1**a**). Interstitial fluid velocity, which is given by Darcy law, is approximately zero at the core of tumors but high at the boundary between tumor and normal tissues (black line in Fig. 1**b**). Both concentration of Doxil and free-doxorubicin in plasma decreased over time after their administrations (figure S4), but the concentration of Doxil decreased more slowly in plasma than free doxorubicin (figure S4). This long-circulating ability of Doxil is because liposomal encapsulation inhibits rapid uptake by the reticulo-endothelial system (RES) and reduces the rate of drug leakage. Additionally, the coating of liposomes with polyethylene-glycol (PEG) confers optimal protection to the vesicles from RES-mediated clearance [52]

*Doxorubicin concentration in tumors after Doxil or doxorubicin administration increased with time, but in a different manner*

Simulated doxorubicin concentration in tumors after administration of Doxil or free doxorubicin alone agreed reasonably well with experimental literature (20) (purple line in figure 2**a** and blue line in figure 2**b**). Note doxorubicin concentration shown here is total amount of doxorubicin in tumors per volume including liposomes in interstitium and cancer cells and released doxorubicin in interstitium and cancer cells. After Doxil administration, doxorubicin concentration in tumors continued to increase, reached maximum at approximately 10 h, and declined gradually (Fig.

2purple line). After free-doxorubicin administration, on the other hand, doxorubicin concentration in tumors also increased; however, it reached maximum at approximately 3 h at approximately 3 μg $ml^{-1}$; it declined steadily afterward (figure 2, blue line). This difference in tumor pharmacokinetics of doxorubicin concentration is due to pharmacokinetics in plasma. Doxil remains longer than free-doxorubicin (figure S4), which allows Doxil to continue to extravasate from blood vessels to tumors. On the other hand, free-doxorubicin disappears from plasma rapidly, which leads to the earlier peak of tumor concentration and decline afterward.

To investigate what factors are more influential in determining the drug concentration in tumors, the volumetric solute flux of convection and diffusion across vessel walls from vessels to tissues, and interstitial fluid flow is analyzed (figure 3**a, c**). Additionally, doxorubicin concentration in tumors is simulated by a different combination of each factor of diffusion across vascular walls, convection through vessel walls, and interstitial fluid flow to investigate what factor is more influential in determining doxorubicin concentration in tumors (figure 6). Figure 3 shows the solute volumetric flux of two factors: i) convection through vessel walls with interstitial flow, and ii) diffusion across vasculature. Note the flux due to convection here is expressed as the net flux of convection through vessel walls and interstitial flow, which is calculated as following:

$$\underbrace{J_F\left(1-\sigma_f\right)c^v}_{\text{convection through vessel walls}} \quad \underbrace{-\frac{\partial}{\partial r}\left(uc\right)}_{\substack{\text{convection due to}\\ \text{interstitial flow}}} \tag{1}$$

because of the following reasons: (i) most of the fluid that comes in tissues from vessel flows out of tissues due to interstitial flow as convection; in a steady state, in which case interstitial fluid pressure does not change over time, the amount of the fluid from vessels is equal to the change in the fluid amount in the interstitium. Additionally, solute flux due to convection is in general large, reaching over 50 μg $ml^{-1}$ $h^{-1}$, for example (at 1 h for Doxil), particularly at the periphery of tumors, and also flux due to interstitial flow is also large there at approximately –40 μg $ml^{-1}$ $h^{-1}$. On the other hand, diffusive flux across vessel walls is less than 1 μg $ml^{-1}$ $h^{-1}$. Thus, calculating convection as the net increase (or decrease) due to both convection through vessel walls and interstitial flow makes it easier to compare it with diffusive flux.

Volumetric convective flux of Doxil in tumors is high at the tumor periphery because of a large pressure difference between capillary and interstitium (solid-line in figure 2a). Diffusive flux across vessel walls is high around $r/R$ = 0.5, except tumor necrotic regions ($r/R$ < 0.4), and it is almost zero at the periphery (figure 3**a**). This smaller diffusive flux across vascular walls at the tumor periphery is because of large *Peclet* number in this area. Following Staverman-Kedem-Katchalsky equation (Eqn. 8), the term included in the diffusion across vascular walls, $Pe/\{\exp(Pe)-1\}$, is smaller when *Peclet* number is larger, Thus, larger *Peclet* number, where convection is more dominant than diffusion (that is, tumor periphery), makes the contribution of diffusion smaller (purple line in figure 3**b**). To see what factor contributes to determining drug concentration in tumors, doxorubicin concentration in tumors is simulated by a different combination of each factor (figure 4). Simulated doxorubicin concentration in tumors due to only diffusion across vessel walls is larger than convection alone (figure 4**a**). However, including diffusion across vascular walls, convection through vessel walls, and interstitial flow leads to higher doxorubicin concentration, which is close to literature value (20) (purple line in figure 4**a**). For free-doxorubicin, on the other hand, contribution of diffusion across vascular walls to the transport across vascular walls is much greater than convection compared with Doxil (blue in figure 3**c,** figure 6a, c) because vascular permeability of BSA with hydrodynamic radius 2-3 nm, almost as large as free-doxorubicin, is approximately six times larger than Doxil with diameter 85 nm (41). This is also clear from the simulated doxorubicin concentration in tumors that only diffusion across vessel walls leads to (figure 6**a**, **c**) much larger doxorubicin concentration than that of convection with interstitial flow (figure 6**a, c**).

*Simultaneous delivery of Clostridium and chemotherapeutic agents in tumors*

Simultaneous delivery of *C. novyi-NT* and Doxil in tumors is simulated to understand the mechanisms of enhanced doxorubicin concentration by the combination of Doxil and *C. novyi-NT*. In order to validate the simulation results obtained in this work against experimental literature work that *C. novyi-NT* was administered 16 h prior to drug administration (20), bacterial transport in tumors is simulated 16 h prior to drug administration, and then drug transport in tumors is also simulated as well as bacterial transport (figure S4). Bacterial concentration in plasma decreased after bacterial administration from $10^8$ – $10^9$ CFU ml$^{-1}$ to $10^7$ CFU ml$^{-1}$ at 40 h post injection (figure S4). Bacterial concentration in tumors agrees reasonably well with literature (42) (53) (figure 5, S5). Bacterial concentration increased rapidly after bacterial administration due to extravasation from blood vessels and reaches at $10^6$ CFU ml$^{-1}$ order immediately after administration. It increased dramatically after 12 h because of bacterial growth (figure 5). The lag period, the duration between the introduction of bacteria and onset of exponential growth, was assumed to be zero because it was not available from literature (53). The relatively higher bacterial concentration in tumors compared with literature during 12–36 h, the exponential phase, indicates that the lag phase period is larger than zero. Additionally, the vascular permeability of *Clostridium* is assumed to be zero due to following reasons. *Clostridium* is rod-shaped with approximately 0.5 µm x1.5 µm. However, the pore cut-off size of colorectal tumor blood vessels, which describes the functional upper limit of the size of a particle that can extravasate from the microvessels, is 400 – 600 nm (54) and no 800-nm microspheres were seen to extravasate from vessels in tumors (55). Thus, we set the vascular permeability of *Clostridium novyi-NT* at zero for simplicity (that is, no passive transport across vascular walls). The possible pathway of bacterial penetration in tumors is at the periphery of tumors where pressure difference is larger, and then entered the tissues across vascular walls due to convection, and migrated in tumors due to motility. Here, using the random motility coefficient of *Salmonella typhimurium* at 5.1 x $10^{-9}$ $m^2$ $s^{-1}$, the mean squared displacement (MSD) of *Salmonella typhimurium* in tumors is calculated at 1900 µm. We assumed that bacteria are spherical with 1 µm diameter and the bacterial migration is inhibited by collagen in the same manner as particles with 1000 nm diameter, following eqn. (20)–(22) (figure S1). Previous experimental work showed *Salmonella* are found in tumor necrotic regions and the average distance between the colonies and the functional vasculature was 750 µm (56). This difference in the penetration distance of bacteria from blood vessels between literature and our modeling is possibly explained by the recent work that revealed that, in a confined disordered porous medium, *E. coli* exhibit not run-and-tumble movement but hopping-and-trapping motility, in which cells perform rapid, directed hops punctuated by intervals of slow, undirected trapping (58). Two-pore models showed the interstitial space with pore radii of 13.8 nm and 1 µm a ratio of 9:1 (57). Thus, the most pores in tumors are inaccessible to bacteria and only a small amount of pores are accessible to bacteria. Consequently, *Salmonella* or *C. novyi-NT* probably exhibit hopping-and-trapping motion in tumors with shorter hopping lengths and longer trapping durations (58). It seems likely that bacteria migrated in tumor necrotic regions and trapped in relatively larger pores than bacteria size, and formed larger colonies due to growth.

This difference in bacterial motility in tumors also probably explains that the penetration rate of *Salmonella* in tumors calculated from the random motility coefficient is much faster than *in vitro* experimental literature (59).

Simulated doxorubicin concentration in tumors for the combination of Doxil and *C. novyi-NT* agreed well with experimental literature data (20) (red-line in figure 2**a**). It is at least twice higher than that of Doxil alone (figure 2**a**), reaching over 10 μg $g^{-1}$ (red solid-line in figure 2**a**). On the other hand, however, the combination of free-doxorubicin and *C. novyi-NT* does not enhance doxorubicin concentration in tumors compared with free-doxorubicin alone (figure 2**b**). The mechanisms of these will be discussed below.

*A mechanistic understanding of enhanced doxorubicin concentration in tumors by co-administration of Doxil and C. novyi-NT*

Figure 6 summarizes how each factor determining drug concentration in tumors is increased or decreased, and thus, influences drug concentration in tumors by bacterial proteolytic activity. The enhanced doxorubicin concentration in tumors by the combination of Doxil and *C. novyi-NT* is due to both reduced interstitial fluid pressure and solid stress alleviation, which are caused by bacterial proteolysis on extracellular matrix of collagen. Since *C. novyi-NT* secrete collagenase that degrades type I, II, and III collagen (45), the collagen content of the tumor interstitium decreased with time by bacterial proteolysis; all the collagen of type I and III is degraded at 16 h (figure not shown). Following the decrease in collagen content of the interstitium, hydraulic conductivity of tumor interstitium increased following eqn. (8). Increased hydraulic conductivity of the interstitium in turn reduces interstitial fluid pressure in tumors (figure 1**a**), which increases convection through vessel walls (figure 4**b**). Reduced interstitial fluid pressure alone increased convection through vessel walls of Doxil greatly did not enhance doxorubicin concentration in tumors (figure **b**). This is because increased hydraulic conductivity in tumor interstitium not only reduces interstitial fluid pressure but increases interstitial fluid velocity (eqn. 1) (figure 1**b**), and thus increases efflux from tissues due to interstitial flow. Consequently, the net volumetric flux of tumor tissues is not increased (Fig. 4b). On the other hand, solid stress alleviation caused by the degradation of extracellular matrix of collagen increases both convection and diffusion across vessel walls in the tumor necrotic regions by increasing vessel density (Fig. 4**a**-**d**, pink in Fig. 6S**b**). Both reduced interstitial fluid pressure and solid stress alleviation increases convection through vessels in tumors (red in figure 3**b**), which thus increases accumulation of Doxil in tumors (Fig. 2, red in Fig. 6S**b**).

On the other hand, concentration of free-doxorubicin in tumors is not enhanced by *C. novyi-NT*, even though extracellular matrix of collagen is degraded by bacterial proteolytic activity (green solid-line in figure 2**b**). The simulation results are consistent with previous experimental literature (green dashed-line in figure 2**b**) (20). This is partly because of larger vascular permeability of free-doxorubicin than Doxil (41). Larger vascular permeability of free-doxorubicin makes the diffusion more dominant rather than convection in determining tumor concentration (Fig. 3**c, d**). This trend can obviously be seen that simulated doxorubicin concentration in tumors by only diffusion across vessel walls (orange in figure 4**c**) is much higher than the simulated one by convection through vessel walls and interstitial flow (light blue in figure 4**c**), compared with Doxil (Fig. 4**a**). On the other hand, reduced interstitial fluid pressure due to the degradation of collagen *decreases* convection (Fig. 3**d**) because increased convection across vessel walls is outweighed by increased efflux due to increased interstitial flow as convection. This different between Doxil and doxorubicin is because of the following reasons: (i) diffusion across vascular walls, not convection through vessel walls, is the major determinant of tumor concentration for free-doxorubicin delivery due to its large vascular permeability, and (ii) diffusive transport is decreased by increased convection through vessel walls following Staverman-Kedem-Katchalsky equation (Eqn. 8) (figure 3**c**, 6S**c**), (iii) increased hydraulic conductivity of interstitium increases interstitial fluid velocity (figure 1**b**), which increases efflux of doxorubicin in interstitium, which is determined by both convection and diffusion because of the reason of (i). Thus, reduced interstitial fluid pressure does not enhance the net transport across vessel walls even though convection through vessel walls is enhanced (figure S6**d**). The simulated doxorubicin concentration in tumors from reduced interstitial fluid pressure alone (gray line in figure 4**d**) is lower than that without bacteria (blue one in figure 6**c**). Solid stress alleviation increases vessel density in tumor necrotic regions and thus, slightly increases doxorubicin concentration in tumors (figure 4**c, d**, pink line in figure S6**d**); this concentration from both reduced interstitial fluid pressure and solid stress alleviation is almost the same as that without *C. novyi-NT* (figure 2**b**, figure S6**d**).

We have also simulated the simultaneous delivery of *Salmonella* and 5-fluorouracil (5-FU) because the combination of engineered therapeutic *Salmonella* and 5-FU led to a prolonged survival rates (14). To follow their experimental procedure, bacteria and 5-FU are administered at day 0 simultaneously. The same parameters for the transport model are used as free-doxorubicin, except for pharmacokinetics in plasma and binding kinetics. Note the colorectal tumor was used in literature work, which makes comparing simulation results in this model with experimental literature results reasonable. The simulated maximum concentration of 5-FU in tumors is at approximately 13 $\mu$g ml$^{-1}$ for both 5-FU alone and 5-FU with bacteria (figure S7). After reaching the maximum, 5-FU concentration in tumors decreased rapidly for 5-FU alone, less than 5% of the maximum concentration at 24 h. The simulated 5-FU concentration in tumors for 5-FU alone agreed reasonably well with experimental literature data (figure S7). On the other hand, the combination of 5-FU and bacteria leads to a slower decline in tumor concentration; still more than 20% of its maximum remains at 24 h. However, the synergetic effect of therapeutic *Salmonella* and 5-FU is not just due to this enhanced drug delivery but also due to the fact that engineered *Salmonella typhimurium* kill cancer cells by releasing encoded anti-cancer agents in tumor necrotic regions, while 5-FU kills the cells close to blood vessels, as discussed previously (18).

*Effect of binding affinity on enhanced drug delivery by bacterial proteolytic activity*

The role of binding affinity in macromolecules transport in tumors is discussed previously (29) (57). Higher binding affinity allows drugs to be taken up more by cancer cells, which makes the interstitium concentration lower. We hypothesize that faster binding affinity leads to larger synergetic effect because faster z binding affinity leads to lower interstitium concentration, which makes the convection due to interstitial flow smaller. When drugs are administered with bacteria because when extracellular matrix of collagen is degraded by bacterial proteolysis, the interstitial fluid velocity is increased (figure 1**b**), thus, less interstitium concentration due to higher binding affinity makes the efflux of interstitial flow smaller, which thus increases drug concentration greater. To validate this, we simulated 5-FU delivery at different binding rates here. Chemotherapeutic agent of 5-FU is chosen because the binding kinetics of 5-FU is described more simply than doxorubicin. Delivery of 5-FU in tumors with or without bacteria is simulated at different binding rate, $k_{21}$. Unexpectedly, faster binding affinity does not enhance the synergetic effect of the combination therapy (figure S7). This is because faster binding affinity makes the interstitium concentration lower, which thus increases the drug concentration in tumors by reducing the efflux due to interstitial flow. However, in the case without bacteria, less interstitium concentration increases diffusion across vascular walls, thus, increases drug concentration in tumors. Therefore, in both case, faster binding rate leads to higher 5-FU concentration in tumors. Note binding affinity does not play a role for the delivery of larger particles such as liposomes, either, because convection through vessel walls is more dominant in the transport across blood vessels, rather than diffusion, (ii) convection is much enhanced by bacterial proteolytic activity compared with smaller molecules (figure 3**d**, 5**d**). This agrees with the previous literature that larger proteins can achieve similar retention to smaller ones with >100-fold weaker binding (57).

We compare the mechanistic understanding obtained in this work with the combination therapy literature, as summarized in table 1. The combination therapy of anti-angiogenesis agents and therapeutic bacteria is discussed not here but in sensitivity analysis part. This understanding (figure 10) is, at least in part, consistent with literature (table 1). Liposomal doxorubicin with diameter 85 nm and trastuzumab, a monoclonal antibody with hydrodynamic radius -20 nm, showed high synergetic efficacy in combination of *C. novyi-NT*. In the combination of therapeutic bacteria and anti-cancer drugs with smaller diameter, some show larger synergy, while the other show the smaller or no synergy. It remains unclear about this reasons. It should be noted that the anti-tumor effect is determined not just vascular permeability, pharmacokinetics, binding affinity, and drug delivery but metabolisms, drug resistance, etc. More importantly, another mechanism for combination therapy is that anti-cancer agent acts on cancer cells in well-vascularized regions, while anaerobic bacteria act on avascular tumor necrotic regions (18). It seems likely that the mechanisms of the combination therapy of therapeutic bacteria and chemotherapy is two-fold: (i) anti-cancer agents kill cancer cells in the areas close to vessels, while therapeutic bacteria kill cells in the areas distal from cells, (ii) delivery of chemotherapeutic agents in tumors with larger diameter is enhanced by bacterial proteolytic activity on extracellular matrix of collagen. This study at least provides an understanding of the mechanisms underlying the combination therapies in previous literature and also a cue to designing an effective chemotherapeutic agent that can be delivered effectively by proteolytic activity of tumor-targeting bacteria.

*Simulated doxorubicin concentration in cancer cells v.s. in vivo anti-tumor efficacy in experimental literature*

To analyze the effect of co-administration of *C. novyi-NT* on anti-cancer effect of Doxil and doxorubicin, the doxorubicin concentration in cancer cells is calculated from the simulations (figure 8). In order for Doxil to be effective, doxorubicin contained in liposomes must be released from Doxil. Additionally, doxorubicin in tumor tissues need to be taken up by cancer cells to exhibit its anti-cancer effect. The process of free-doxorubicin uptake by cancer cells after Doxil administration contains two pathways: (i) doxorubicin is released from liposomes in the interstitium, and then the released one is taken up by cancer cells, and (ii) Doxil is taken up by cancer cells, and then the doxorubicin contained in liposomes is released there. For free-doxorubicin, doxorubicin in the interstitium needs to be taken up by cancer cells. The case for free-doxorubicin is discussed here first due to its simplicity of the process. Doxorubicin concentration in cancer cell for free-doxorubicin increased immediately after drug administration, irrespective of bacteria (figure 8**c**); however, it reached at its maximum at approximately 6 h, and then decreased gradually. This is because doxorubicin concentration in cancer cells reaches an equilibrium concentration with extracellular concentration. Note the total concentration in tumors shown in this figure is calculated as: $c_f \phi_f + c_c \phi_{cell}$; the actual extracellular (interstitium) concentration is higher than the value in this figure. On the other hand, for Doxil, the concentration of Doxil in the interstitium decreases gradually because it is taken up by cancer cells and it releases liposome-containing doxorubicin. Finally, doxorubicin in cancer cells continues to increase due to both the release from Doxil and uptake of doxorubicin in the interstitium by cancer cells (figure 8**a**). Consequently, the simulated pharmacokinetics of doxorubicin in cancer cells for Doxil treatment is different from free-doxorubicin. To be specific, since free-doxorubicin does not have a process of liposome-release, doxorubicin concentration in cancer cells for free-doxorubicin increases more rapidly than Doxil. However, after reaching its maximum, the doxorubicin in cancer cells decreases gradually because of a rapid decay in plasma concentration (figure S4). On the other hand, for Doxil, the doxorubicin concentration in cancer cells increases more slowly than free-doxorubicin (figure 8**a**), however, it continues to increase because of the long-circulating ability of Doxil in plasma (44), reaching approximately 1.5 μg ml$^{-1}$ at 120 h post-injection. The higher simulated doxorubicin concentration in cancer cells for Doxil than doxorubicin, at the same dose, implies that Doxil is more effective than doxorubicin on cancer cells, even though Doxil has a process to release encapsulated doxorubicin. By administrating bacteria together with Doxil, the total doxorubicin concentration in tumors is over twice larger with bacteria than that without bacteria (figure 8b). Thus, by the combination of Doxil and bacteria, doxorubicin concentration in cancer cells is over twice increased compared with Doxil alone, reaching at 3.5 μg ml$^{-1}$ at 120 h post-injection. These simulated doxorubicin concentration in cancer cells are consistent with the previous experimental literature that the anti-tumor effect on CT26 or HCT116 colorectal cancer by Doxil is slightly larger than free-doxorubicin (20). More importantly, they showed that though, all mice of doxorubicin and Doxil treatment died by 30 days and 45 days, respectively, the combination of Doxil and *C. novyi-NT* spores resulted in complete regression of tumors in 100% of mice and 65% of the mice were still alive at 90 days. Our modeling results agree with these experimental results because the simulated doxorubicin concentration in cancer cells by Doxil with bacteria is at least twice higher than Doxil without bacteria (20). The simulated doxorubicin concentration in cancer cells after co-administration of free-doxorubicin and bacteria is almost the same as that of free-doxorubicin alone (Fig.8**c**). 100% of the mice treated with *C. novyi-NT* and free doxorubicin at the same dose died within 2 weeks, while 100% of the untreated or doxorubicin-treated mice died by 20 days or 30 days, respectively (20). Our modeling results are at least consistent with the experimental results in that the combination of free-doxorubicin and *C. novyi-NT* does not enhance doxorubicin concentration in tumors nor improve chemotherapeutic efficacy.

It is worth noting here that heat-inactivated *C. noyvi-NT* did not enhance the anti-tumor effect of Doxil (20). The reason of this is probably because heat-inactivated bacteria lose motility (60); thus, they are not able to penetrate in tumors. Random motility coefficient of immotile bacteria is approximately two or three order smaller than that of motile bacteria. The random motility coefficient of immotile *Salmonella typhimurium* in water at 4.3 x $10^{-13}$ $m^2$ $s^{-1}$ (61). The calculated mean squared displacement (MSD) in tumors at 16 h is 42.0 μm, without considering the effect of collagen inhibition on bacterial migration in tumors (that is, all tumor interstitium collagen assumed to be degraded). Actual distance of bacterial migration from blood vessels when including collagen inhibition on immotile bacterial migration should be much smaller. It seems that the inhibition of collagen on movement of immotile bacteria is greater because diffusion of particles with larger diameter are more inhibited by extracellular matrix of collagen (figure S2). As a result, heat-inactivated bacteria, which is no longer motile, are not able to penetrate in tumors. Therefore, even though heat-inactivated bacteria keep their proteolytic activity, as described in previous work (35), little collagen in tumors is degraded by bacterial proteolysis due to their poor penetration nor drug delivery is enhanced.

*Parametric sensitivity analysis*

Parametric sensitivity analysis is carried out to investigate which factor is influential in determining doxorubicin concentration in tumors. Each parameter is changed +30% or –30%, for example, and then how much the simulated doxorubicin concentration in tumors is changed following the increase or decrease in each parameter is analyzed. Vascular permeability of Doxil in tumor vessels is sensitive in determining doxorubicin concentration in tumors without bacteria, though it is not as sensitive when *C. novyi-NT* is included (figure 9). This implicates that diffusion across vascular walls is dominant for Doxil alone, but when *C. novyi-NT* is included, not diffusion but convection is dominant in the transport of Doxil across vessel walls. This is also supported from the analysis results that doxorubicin concentration in tumors is sensitive to solvent drag reflection coefficient with bacteria, a determinant of convection through vessel walls, while it is not as sensitive for Doxil alone (Fig. 9).  On the other hand, doxorubicin concentration is sensitive to vascular permeability for both with and without *C. novyi-NT* (Fig. S10). This is consistent with the previous discussions about size difference between Doxil and free-doxorubicin and the factors determining doxorubicin concentration in tumors. Diffusivity is not sensitive at all for both Doxil and doxorubicin, irrespective of addition of *C. novyi-NT*. Doxorubicin concentration in tumors for Doxil with *C. novyi-NT* is also sensitive to proteolysis rate and vascular density. Note that doxorubicin concentration in tumors does not change when proteolysis rate is increased because all the collagen is degraded at 16 h after bacterial injection, when Doxil or free-doxorubicin is administered. Dose and rate constant of pharmacokinetics in plasma affect doxorubicin concentration to the same extent.

Notably, Darcy hydraulic conductivity of interstitium and Starling hydraulic conductivity of blood vessel walls are not sensitive for free-doxorubicin, irrespective of *C. novyi-NT* (Fig. S10) or Doxil alone. But they are much more sensitive for Doxil in combination of *C. novyi-NT* (figure 9). This is explained by a previous discussions about hydraulic conductivity and interstitial conductivity and interstitial fluid pressure (30). The decrease in interstitial fluid pressure by changing Darcy or Starling hydraulic conductivity is dependent on the factor, $\alpha = \sqrt{SL_p/VK}$ , where $L_p$ is Starling hydraulic conductivity, $K$ is Darcy interstitium hydraulic conductivity, $S/V$ is surface area per volume (also see eqn. 6 in supporting information). When $\alpha$ decreases from 50 to 10, decrease in interstitial fluid pressure in tumors is rather small. However, when $\alpha$ decreases from 10 to 1, interstitial fluid pressure decreases greatly (30). Thus, for the administration of *C. novyi-NT* with Doxil, hydraulic conductivity of the interstitium is already decreased by bacterial proteolysis on collagen (eqn. 6), and thus, interstitial fluid pressure decreases greatly with the reduced vascular hydraulic conductivity, $L_p$, caused by vascular normalization by anti-angiogenesis agent of endostatin (62) or bevacizumab (56) (63). Thus, doxorubicin concentration in tumors is easier to be increased when hydraulic conductivity of interstitium or vascular walls is increased (for $K$) or decreased (for $L_p$). In the modeling of this study, $\alpha$ is at 29.3 at the initial period of treatment (or without bacteria), but $\alpha$ decreased due to bacterial proteolysis and is at nine at 16 h post bacterial administration. Thus, the doxorubicin concentraton is very sensitive to hydraulic conductivity. These sensitivity analysis with discussions here also provides a mechanistic understanding of the synergetic effect of combination of *Salmonella typhimurium* and angiogenesis inhibitor, endostatin (23) or *bevacizumab* (25). Treatment of *bevacizumab* (BEV) and *gemicitabine* (GEM) followed by *S. typhimurium A1-R* significantly reduced tumor weight compared to BEV/GEM treatment alone (25). The mechanisms of this is probably because the combination of *Salmonella* and anti-angiogenesis agents dramatically reduced interstitial fluid pressure because of increased hydraulic conductivity in interstitium by degradation of extracellular matrix of collagen by bacterial proteolysis and normalized tumor vasculature and reduced hydraulic conductivity of tumor vasculature by *bevacizumab* (33). Though this reduced interstitial fluid pressure in tumors needs to be validated by *in vivo* experiment, it increases drug concentration in tumors by increasing convection of drugs through vessel walls greatly.

To summarize, a mechanistic understanding of enhanced concentration of chemotherapeutic agents in tumors is provided (figure 10). Simulated doxorubicin concentration in tumors is enhanced by the combination of Doxil and *C. novyi-NT* than Doxil alone due to both reduced interstitial fluid pressure and solid stress alleviation, caused by the degradation of extracellular matrix of collagen by bacterial proteolytic activity. Degradation of extracellular matrix of collagen increases hydraulic conductivity of interstitium, which reduces interstitial fluid pressure in tumors, while it leads to solid stress alleviation, which increases vessel density in tumor necrotic regions by decompressing blood vessels. The simulated doxorubicin concentration in cancer cells by the combination of Doxil and *C. novyi-NT* is higher than that of Doxil alone, which agreed with enhanced anti-tumor efficacy by this combination in previous experimental literature *in vivo.* Chemotherapeutic agents with lower vascular permeability (larger diameter) or faster uptake by cancer cells have greater potential for their deliveries to be improved in combination with bacterial cancer therapy. Sensitivity analysis along with previous literature shows that the anti-angiogenesis strategy can be used to further enhance this combination therapy by reducing interstitial fluid pressure greatly.

3. Methods

Interstitial fluid flow, drug transport, and bacterial transport in tumors are mathematically modeled. Interstitial fluid flow in tumor tissues is described by Darcy's law as:

$$u = -K \frac{\partial p}{\partial r} \tag{2}$$

The governing equation of interstitial fluid transport is obtained by following Starling's law as below:

$$K \frac{\partial^2 p}{\partial r^2} = \underbrace{\Phi_V}_{\substack{\text{Flow from} \\ \text{blood} \\ \text{vessels}}} + \underbrace{\Phi_L}_{\substack{\text{Drainage to} \\ \text{lymphatic} \\ \text{vessels}}} \tag{3}$$

$$\Phi_V = \frac{J_V}{V} = \frac{L_p S}{V} \underbrace{\left\{ \left( p_v - p \right) - \sigma \left( \pi_v - \pi_i \right) \right\}}_{\text{net driving force}} \tag{4}$$

$$\Phi_L = \frac{J_L}{V} = \frac{L_{pL} S_L}{V} \left\{ p_L - p - \sigma_L \left( \pi_L - \pi_i \right) \right\} \tag{5}$$

Hydraulic conductivity in tissues negatively correlated with collagen content and glycosaminoglycan in the interstitium with determination coefficient $R^2$ of 0.86 and 0.80, respectively (figure S1). Thus, the Darcy hydraulic conductivity of the interstitium of the tumors is described with collagen content as:

$$\log_{10}(\frac{K}{K_0}) = -1.85\log_{10}(\frac{c_c}{c_{c0}}) \tag{6}$$

The coefficient is determined from previous literature [5] (36) [65] (figure S1).

Transport of chemotherapeutic agents in tumors

The governing equation of transport of Doxil in tumors is described as:

$$\frac{\partial\left(c^i\phi_l\right)}{\partial t} = \underbrace{\overset{\substack{\text{effective}\\ \text{diffusion}\\ \text{coefficient}}}{D_{eff}}\frac{\partial^2\left(c^i\phi_l\right)}{\partial r^2}}_{\text{diffusion}} \underbrace{-\frac{\partial}{\partial r}(\overset{\substack{\text{interstitial fluid}\\ \text{velocity}}}{u}\; c^i)}_{\text{interstitial fluid flow}} + \underbrace{\overset{\substack{\text{Solute flux}\\ \text{across vascular walls}}}{J_s}\;\frac{S}{V}}_{\substack{\text{extravasation from}\\ \text{blood vessels}}} \underbrace{-\overbrace{k_{lip}^{uptake}}^{\substack{\text{uptake rate}\\ \text{by cancer cells}}}\left(c^i\phi_l\right)}_{\substack{\text{uptake to}\\ \text{cancer cells}}} \underbrace{-\overbrace{k_{rel}^i}^{\substack{\text{release rate}\\ \text{from liposomes}\\ \text{in interstitum}}}\left(c^i\phi_l\right)}_{\text{release from liposomes}} \tag{7}$$

Transport of drugs across vasculature, which is described in the term, $J_s$, and the solute flux, $J_s$ [mol m$^{-2}$ s$^{-1}$] from blood vessels is given by Staverman-Kedem-Katchalsky equation:

$$J_s = \underbrace{J_F(1-\sigma_F)c_v}_{\substack{\text{convection}\\ \text{through}\\ \text{blood vessel walls}}} + \underbrace{P(c_v - c\,\phi)\frac{Pe}{\exp(Pe)-1}}_{\text{transvascular diffusion}} \tag{8}$$

The governing equation of bacterial transport in tumors is described as:

$$\frac{\partial\left(b\varphi_b^t\right)}{\partial t} = \underbrace{\mu_{eff}\frac{\partial^2\left(b\varphi_b^t\right)}{\partial r^2}}_{\text{Motility}} \underbrace{-\frac{\partial}{\partial r}\{ub\}}_{\text{Interstitial flow}} + \underbrace{J_F(1-\sigma_F)b}_{\substack{\text{convection}\\ \text{through}\\ \text{blood vessel walls}}} + \underbrace{P(b_v - b\,\phi)\frac{Pe}{\exp(Pe)-1}}_{\text{transvascular diffusion}} + \underbrace{\frac{q(t)}{1+q(t)}u_{\max}\left(1-\left(\frac{b\phi}{b_{\max}}\right)\right)b\phi}_{\text{Growth}} \tag{9}$$

Proteolysis on extracellular matrix of collagen by *C. novyi-NT* is included in the modeling and proteolysis reaction rate is expressed in Michalis-Menten kinetics as following:

$$\frac{\partial c_c}{\partial t} = \frac{V_{max}c_c}{K_m + c_c} \tag{10}$$

where $V_{max}$ [μg ml$^{-1}$ min$^{-1}$] is maximum reaction rate and $K_m$ [μg ml$^{-1}$] is Michaelis constant. *C.*

*novyi-NT* secrete collagenase *ColG,* class I, M9B family, which degrades collagen type I, II, and III (66). *C. novyi-NT* do not secrete hyaluronidase (45).

The role of solid stress on blood vessel density is included in this model for simplicity as below:, since all degradable collagen of type I, II, III is degraded by bacterial proteolysis before drug injection, the surface area of blood vessel per volume is changed as following.

$$\frac{S}{V} = 2000 \ (r/R < 0.4) \text{ (without bacteria)} \tag{31}$$

$$\frac{S}{V} = 12000 \ (r/R < 0.4) \text{ (with bacteria)} \tag{32}$$

Acknowledgements

This research is financially supported by the Keio Gijuku Academic Development Funds.

Discussions with Prof. Tsuyoshi Osawa, Yukiko Matsunaga, Toru Torii, and Eriko Yasunaga of the University of Tokyo are greatly appreciated. Discussions with Prof. Datta of Cornell University and Dr. Filippo Menolascina of the University of Edinburgh are also appreciated greatly.

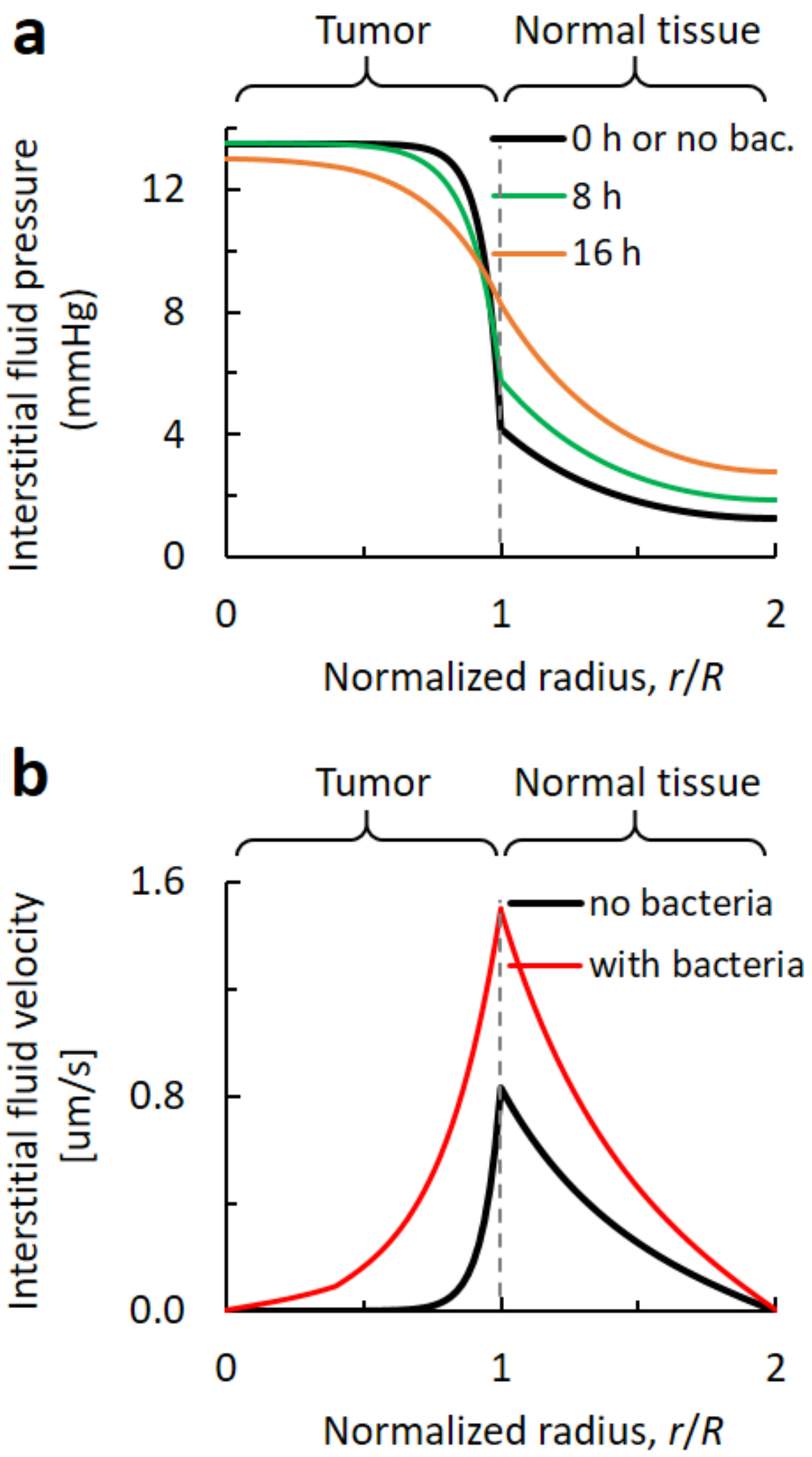

Figure 1. Interstitial fluid pressure in tumors decreased by bacterial proteolytic activity because it increases hydraulic conductivity by degrading extracellular matrix of collagen (green and orange lines in **a**). Interstitial fluid velocity is increased by bacterial proteolytic activity because it degrades extracellular matrix of collagen, which increases hydraulic conductivity of interstitium (red line in **b**). Time indicated in **a** shows the elapsed hours after bacterial injection.

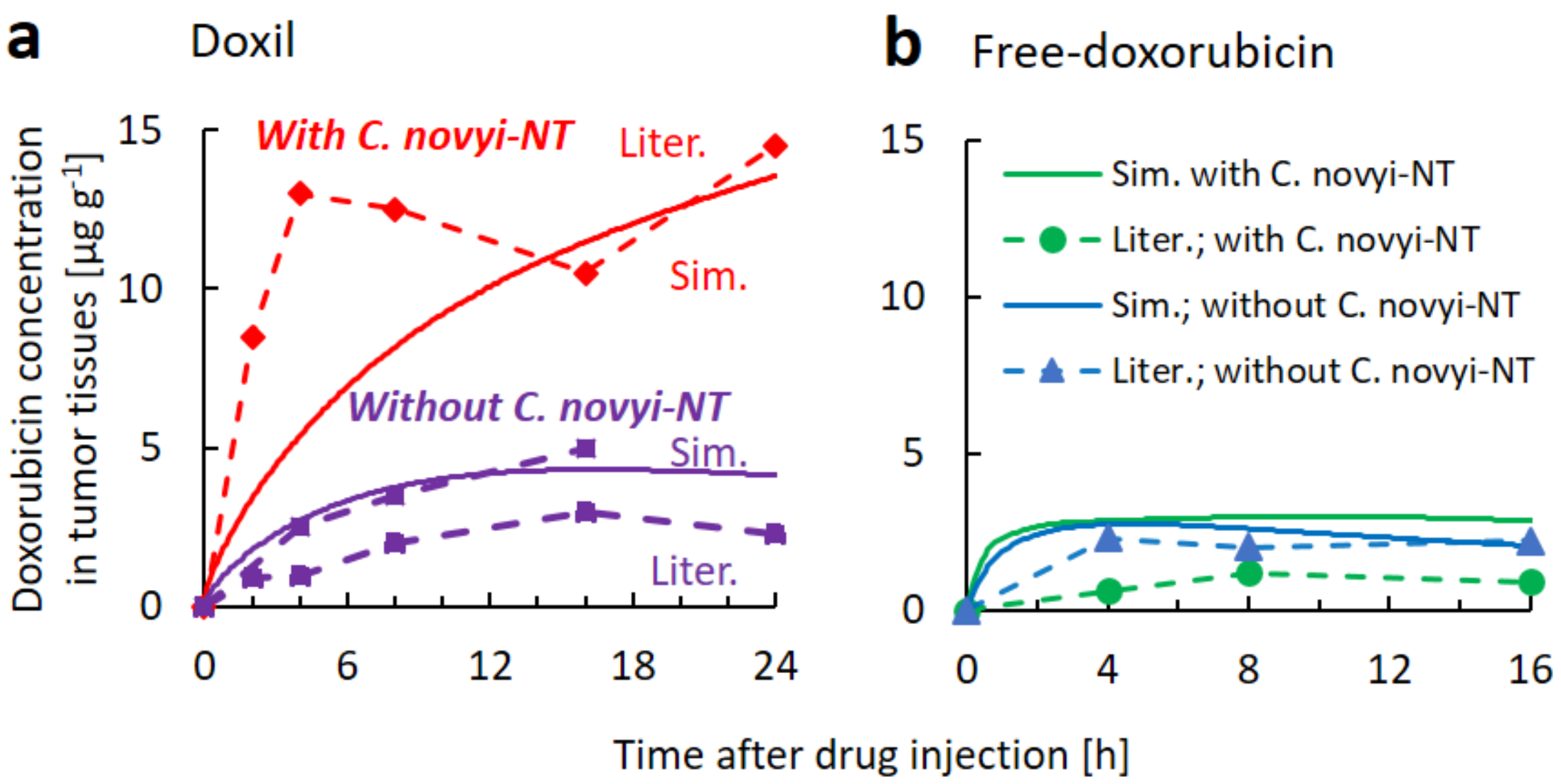

Figure 2. Doxorubicin concentration in tumors of Doxil treatment is enhanced with *C. novy-NT* (red solid-line) compared with that without *C. novyi-NT* (purple solid-line) (**a**), though doxorubicin concentration in tumors of free-doxorubicin treatment is not enhanced by *C. novyi-NT* (**b**). Solid-line: simulation results; dashed-line with plots: literature data (23). The literature data of Doxil without *C. novyi-NT* includes those from main article (bottom ones) and supplementary information (top ones).

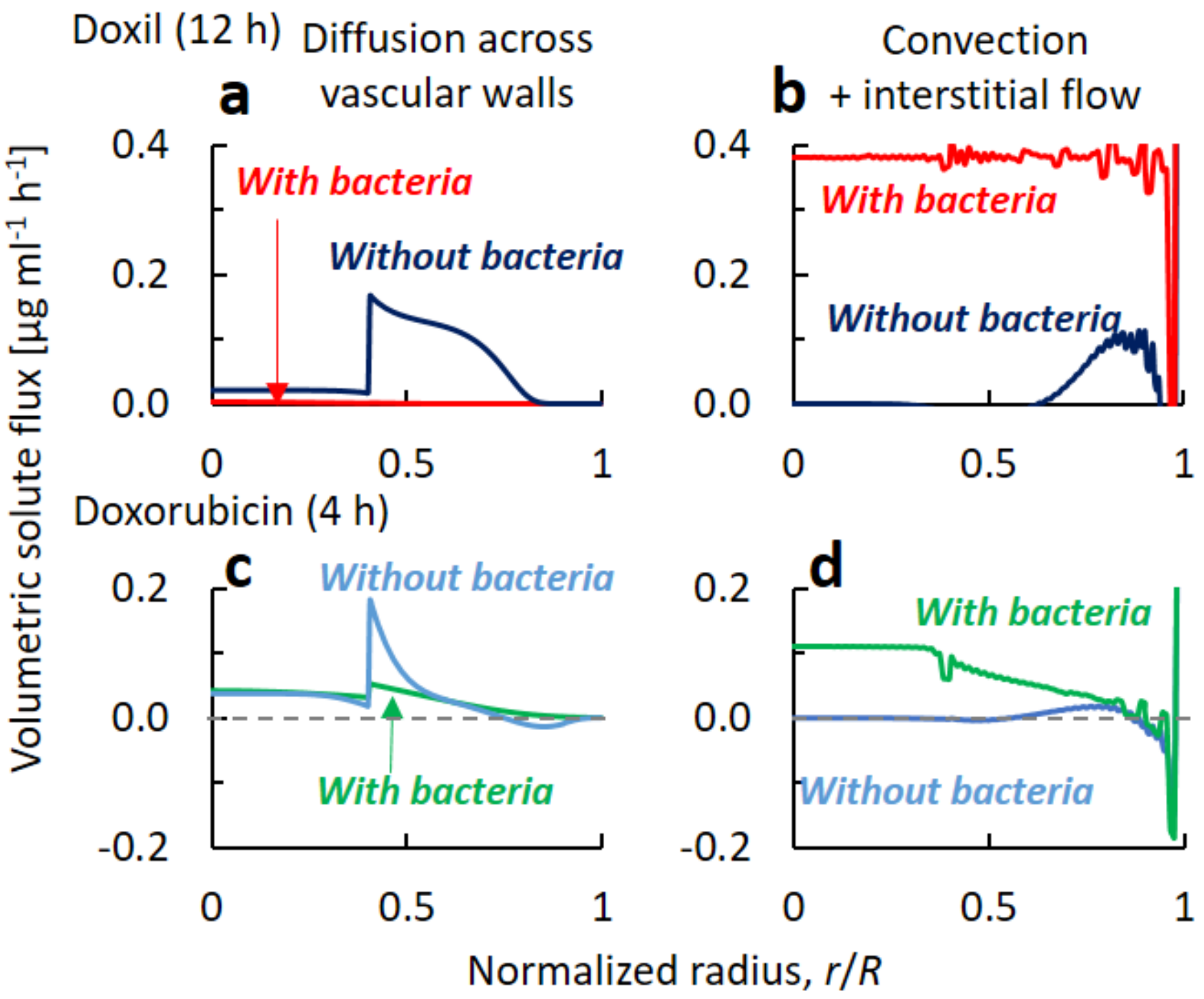

Figure 3. Though bacterial proteolytic activity decreases diffusion across vascular walls (a, c), it enhances the net flux of convection through vessel walls and that due to interstitial flow (b, d). Convective transport of Doxil is relatively large (deep blue in **a**) compared with diffusive transport (**b**) because of its smaller vascular permeability of tumor vessels, while diffusive transport is dominant over convection for free-doxorubicin (**c, d**).

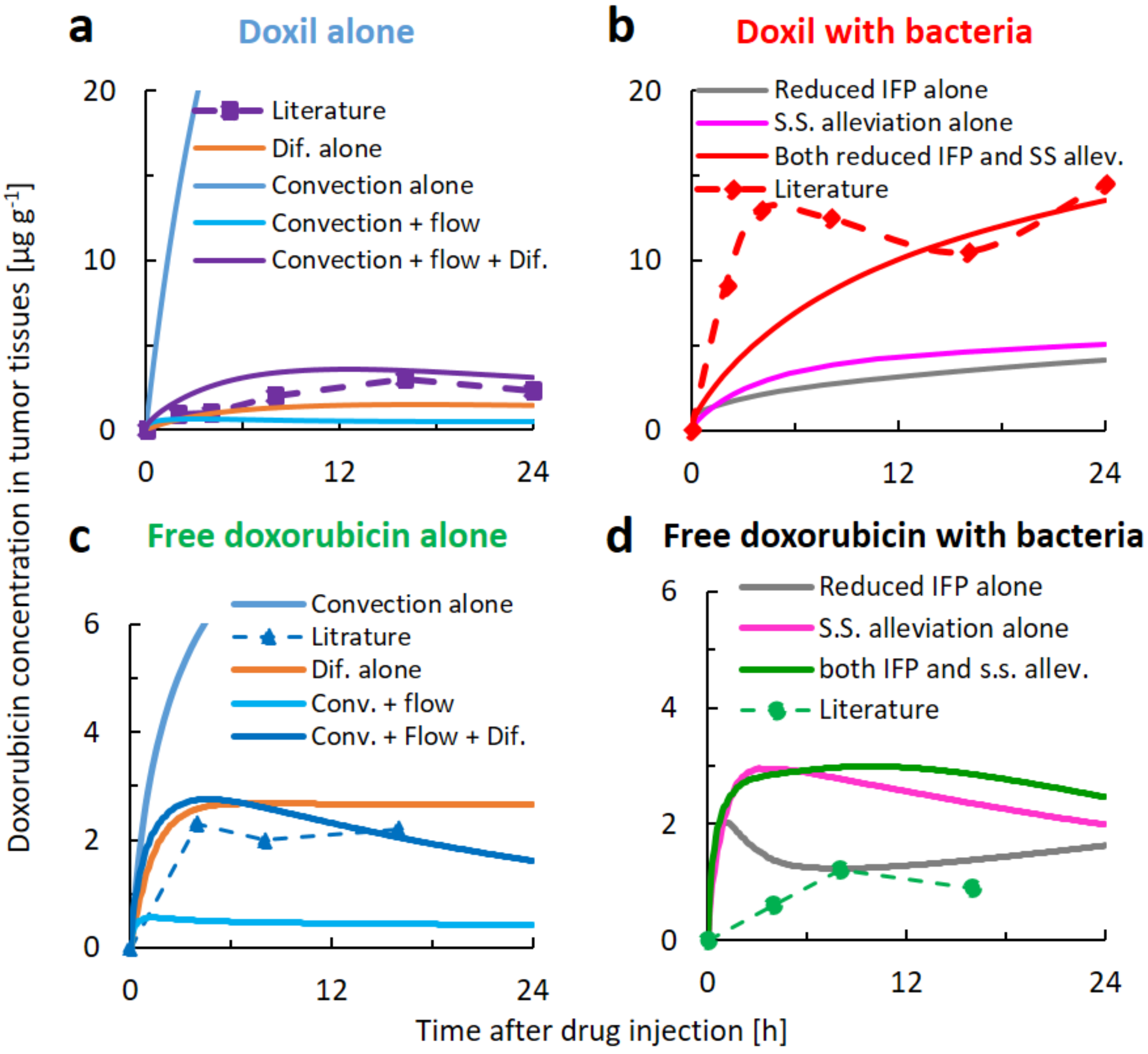

Figure 4. Simulated doxorubicin concentration in tumors by different combinations of each factor such as diffusion across vascular walls, convection through vessel walls, and interstitial flow in the transport of Doxil alone (**a**), Doxil with bacteria (**b**), free-doxorubicin alone (**c**), and free-doxorubicin with bacteria (**d**). Dashed lines with plots are measured doxorubicin concentration in previous literature (20).

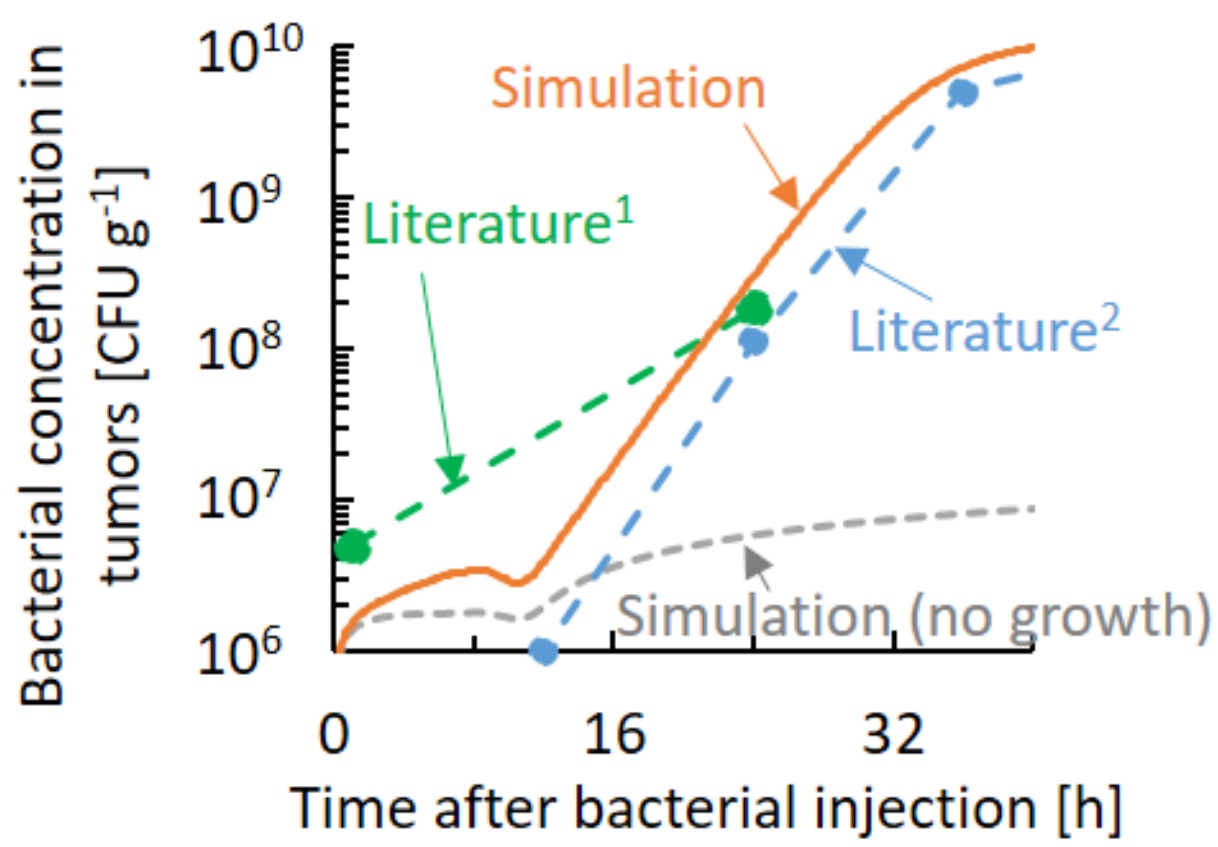

Figure 5. Simulated bacterial concentration in tumors reaches $10^6$ – $10^7$ CFU $ml^{-1}$ immediately after bacterial injection (orange line) due to extravasation from vessels and continued to increase gradually. It increases rapidly after 12 h due to bacterial growth, reaching over $10^9$ CFU $ml^{-1}$. Solid-line: simulation results; dashed-line: literature (green: (53); blue: (42)). Gray dotted-line: simulation results without bacterial growth. Bacterial concentration is shown in logarithmic scale. Shrinkage of this figure is shown in supporting information.

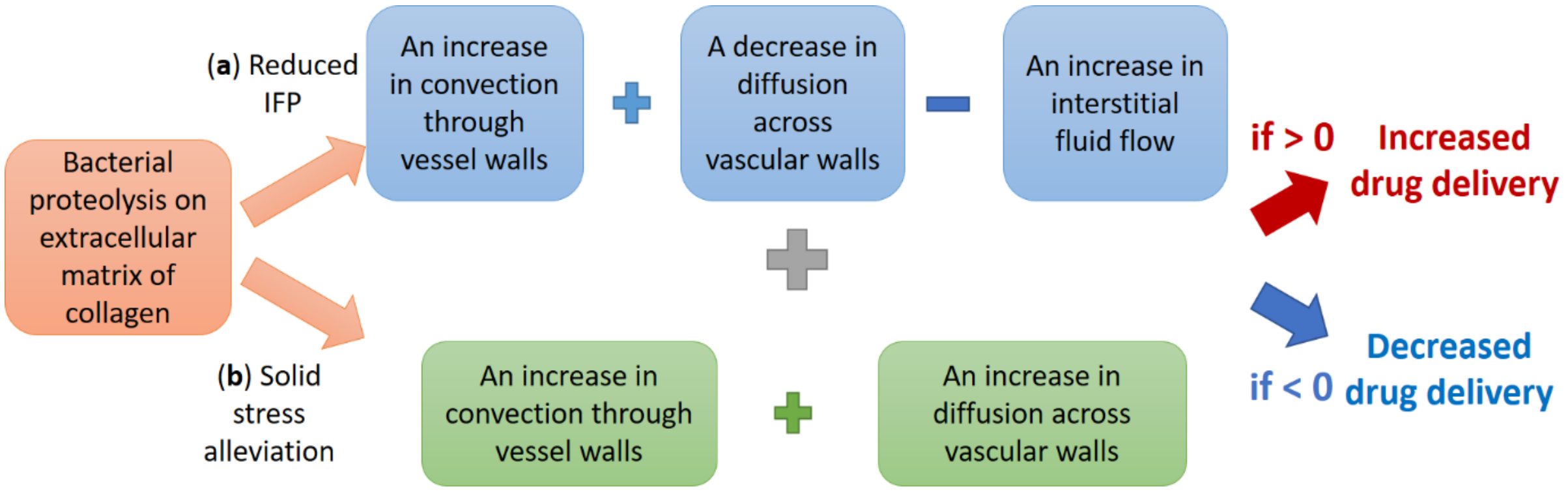

Figure 6. Bacterial proteolysis on collagen influences delivery of chemotherapeutic agents in tumors via two pathways: (**a**) reduced interstitial fluid pressure by collagen degradation increases convection through vessel walls, decreases diffusion across vascular walls, and increases interstitial fluid flow, (**b**) solid stress alleviation of tumors increases both convection through vessel walls and diffusion across vascular walls. The net increase in solute flux of these transports leads to increased drug delivery, though net decrease leads to decreased drug delivery.

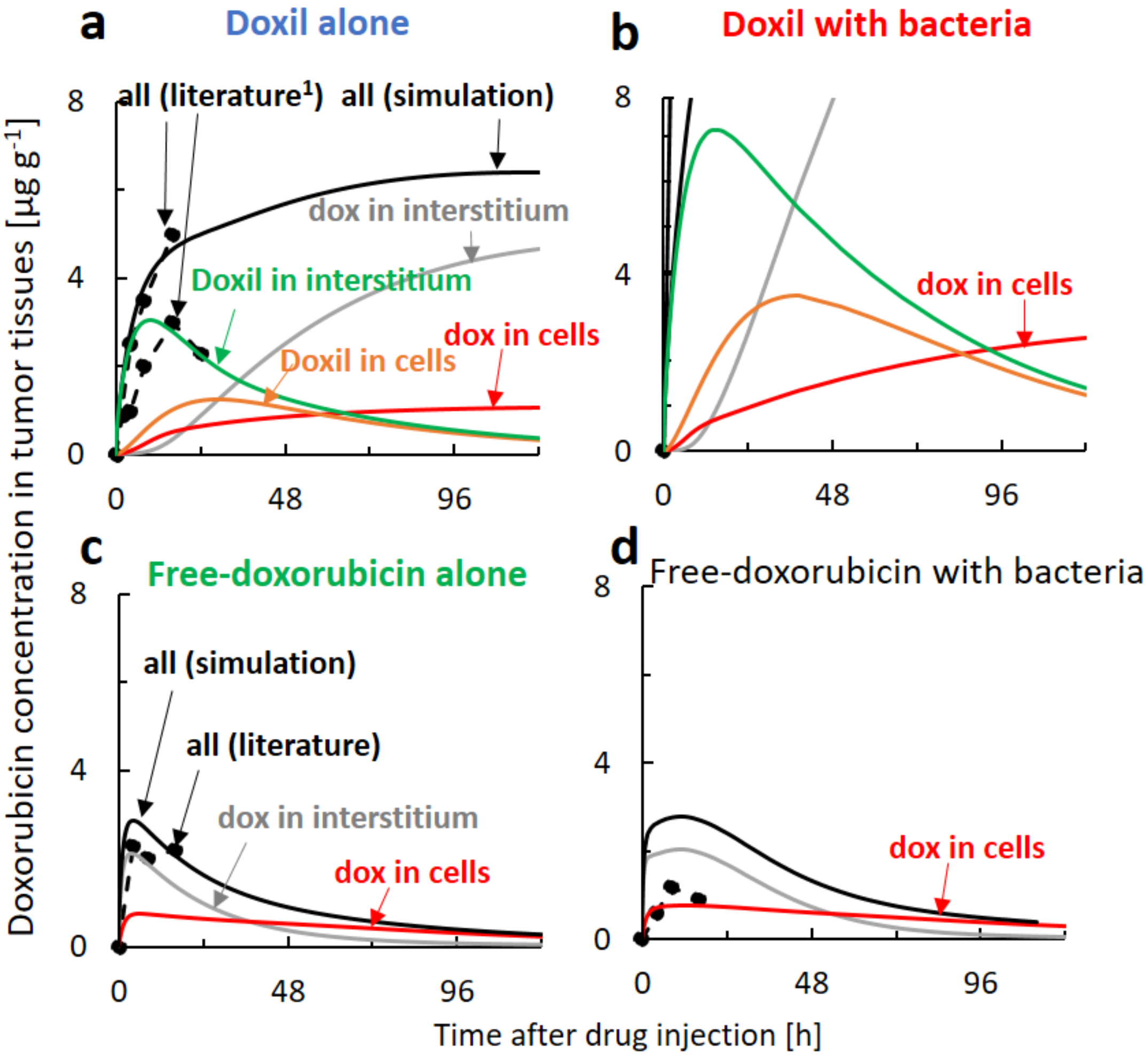

Figure 8. Simulated concentration of all doxorubicin in tumors (black), Doxil in interstitium (green), Doxil in cancer cells (orange), doxorubicin in the interstitium (gray), and in cancer cells (red) for Doxil without bacteria (**a**), Doxil with bacteria (**b**), free-doxorubicin without bacteria (**c**), and free-doxorubicin with bacteria (**d**). Dashed-line in black is measured doxorubicin concentration *in vivo* in literature (20). Note the literature data in (**a**) includes both data in main article and supporting information (20). dox: free-doxorubicin. The figure of **b** with enlarged vertical scale is in supporting information.

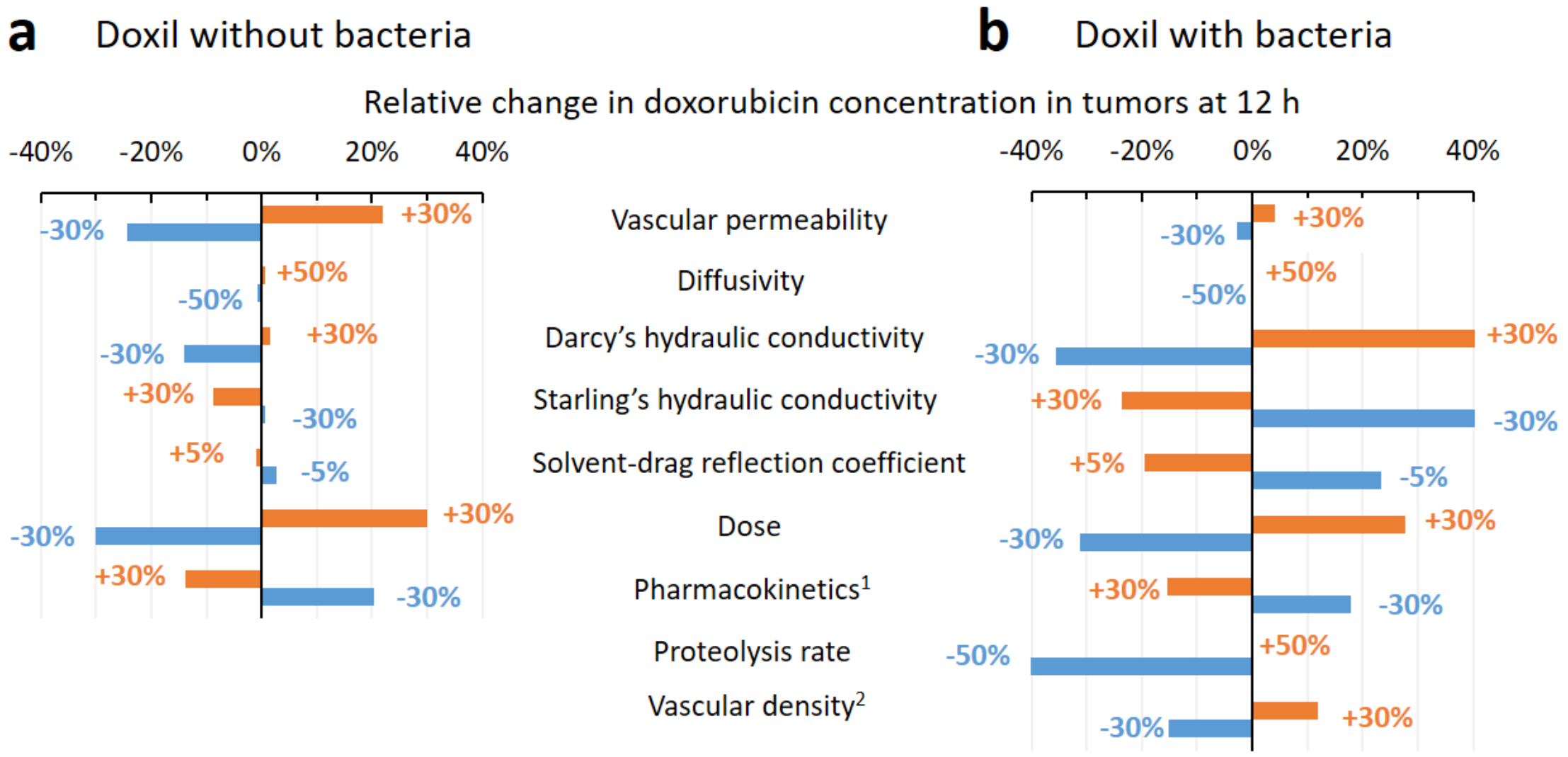

Figure 9. Darcy and Starling's hydraulic conductivity are not sensitive to determining doxorubicin concentration in tumors for Doxil alone but very sensitive for Doxil with bacteria. Parametric sensitivity analysis for determining drug concentration in tumors at 12 h of Doxil alone (**a**), Doxil with *C. novyi-NT* (**b**).

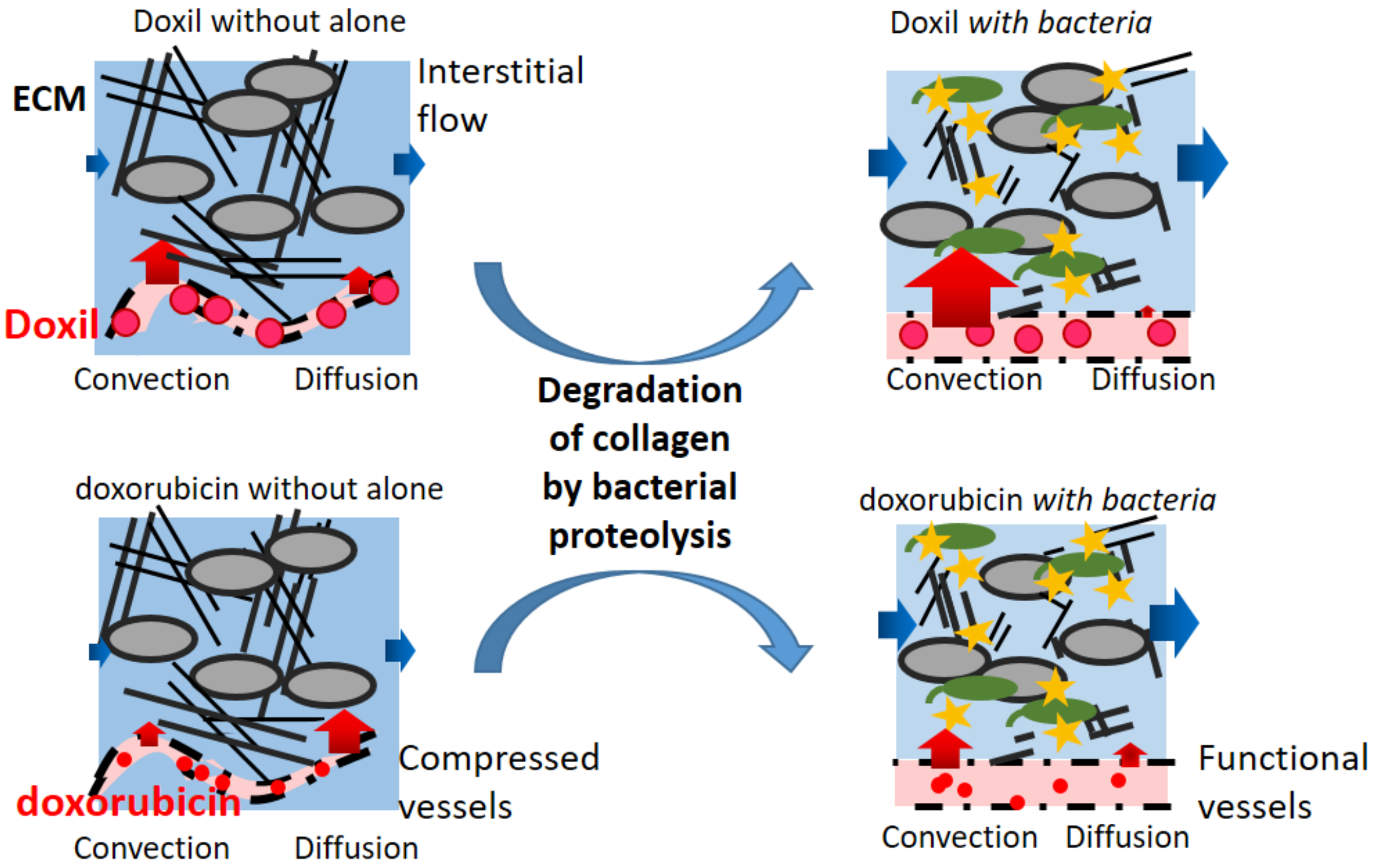

Figure 10. A summary of a mechanistic understanding provided in this work. Degradation of extracellular matrix of collagen by bacterial proteolysis increases convection of Doxil through vessel walls by reduced interstitial fluid pressure and solid stress alleviation (a, b), though it does not enhance doxorubicin concentration for free-doxorubicin treatment (c, d) because larger vascular permeability of free-doxorubicin allows diffusion more dominant rather than convection and increased hydraulic conductivity increases interstitial fluid velocity, and thus, increases efflux of drugs from tissues.

Table 1. List of previous experimental work of combination of chemotherapy and bacterial cancer therapy

| Bacteria | Genetic engineering of bacteria | Drug | Drug type | $T_{1/2}$ [h] | Protein binding | Synergetic effect | Cancer cell | Reference |
|---|---|---|---|---|---|---|---|---|
| *C. novyi-NT* | | Dolstatin-10 (D-10) + mytomycin C (MMC) | Cytotoxic (D-10); MMC: anti-tumor antibiotic | D-10: $t_{1/2}\ \alpha$ 0.04[2], $t_{1/2}\ \beta$ : 1.6 ; MMC: 0.81 | D-10: 95%[3]; MMC: 24% | Large | colorectal cancer | (17) |
| | | Vinorelbine | Anti-microtubule | 21.4 | 89% | Large | colorectal cancer | (18) |
| | | Docetaxel | Anti-microtubule | 41 | 97% Kd = 10 nmol l–1 | Small | colorectal cancer | (18) |
| | | MAC321 | Analogue of docetaxel | | ~95% | Small | colorectal cancer | (18) |
| | | Paclitaxel | Anti-microtubule | 20.2 | 89-98% | Small | colorectal cancer | (18) |
| | | Vinblastine | Anti-microtubule | 26.2 | 98-99% | No | colorectal cancer | (18) |
| | | Vincristine | Anti-microtubule | - | 75% | Small | colorectal cancer | (18) |
| | | Non-lipoomal doxorubicin | Interacts with DNA and inhibit macromolecule production | 14.2 | 75% | No | colorectal cancer | (20) |
| | | Liposomal doxorubicin (Doxil) | PEGylated liposomal doxorubicin | 55 | 70% | Large | colorectal cancer | (20) |
| *Salmonella choleraesuis* | | cisplatin | interferes with DNA replication | 0.44 | * | Large | Murine lung tumor and hepatoma | (22) |
| *Salmonella typhimurium VNP20009* | | endostatin | anti-angiogenesis | | | Large | Murine melanoma | (23) |
| *Salmonella typhimurium VNP20009* | | Cyclophosphamide (261 g/mol) | Its metabolite phosphoramide mustard forms DNA crosslinks | 3-12 | 20% | Large | Murine melanoma | (24) |
| *Salmonella typhimurium A1-R*[4] | | gemcitabine (GEM), bevacizumab (BEV) | GEM: replaces building blocks of nucleic acids BEV: anti-angiogenesis | 0.23 (GEM) | Negligi-ble (GEM) | Large | Pancreatic cancer | (25) |
| *Salmonella typhimurium A1-R* | | Trastuzumab (-148 kDa) | a recombinant IgG1 $\kappa$ | 12 days | | Large | HER-2-Positive Cervical Cancer | (26) |
| *Salmonella typhimurium* | Secrete Chemokine, Haemolysin, and Pro-apoptotic peptide | 5-fluorouracil | inhibits DNA synthesis | | 10% | Large | MC26 colorectal metastasis in liver | (14) |
| *Salmonella typhimurium VNP20009* | Carry scrambled *shRNA* or *Sox2 shRNA* construct | HM-3 | angi-angiogenesis polypeptide | | | Large | A549 lung cancer | (27) |

| | | | | | | | |
|---|---|---|---|---|---|---|---|
| *Salmonella typhimurium A1-R* | cisplatin and r*MET*ase | Interferes with DNA replication | 0.44 | * | Large | cisplatinum-resistant metastatic osteosarcoma | (28) |

$T_{1/2}$ and protein binding are obtained from Liston et al. (2017) (42).
2. Aherne et al. (1995) (43)
3. De Jonge et al. (2005) (44)
4. *Salmonella typhimurium A1-R* is auxotrophic for *leu* and *arg* which attenuates bacterial growth in normal tissue but allows high tumor virulence (25)
rMETase: recombinant methionidase

Table 2. List of anaerobic bacteria that have been investigated for bacterial cancer therapy

| Bactreia (strain) | Aerobic or anaerobic[1] | Motility | Collagenase | Collagenase family, class, name[2] | Degradable collagen type | Reference about collagenase | Litera comb thera |
|---|---|---|---|---|---|---|---|
| *Clostridium novyi-NT* | – – | + | + | M9B, class I, *ColG* | I, II, III | (45) | (17) (20) |
| *Salmonella typhimurium* (*S. choleraesuis*) | – | + | + | U32 | I | (46) | (14) ( |
| *Clostridium acetobutylicum* | – – | + | + | U32 | I | (47) | |
| *Clostridium sporegenes* | – – | + | + | M9B, class I, *ColG* | I, II, III, (IV) | (48) | |
| *Bifidobacterium longum* | – – | – | – | | | | |

[1] –: facultative anaerobic; – –: obligate anaerobic

[2] *Collagenase name, class, and family are obtained from* https://www.uniprot.org/. (49)

Supporting information

Methods

## 2. Mathematical modeling

Interstitial fluid flow, transport of chemotherapeutic agents, and bacterial transport in tumors are mathematically modeled. The details of these models are to be described here. The parameters used in the modeling will be described at the end of this section.

### 2.1 Geometry

Three-dimensional cylindroids consisting of tumor tissues ($0 < r/R < 1$) and normal tissues ($1 < r/R < 2$) are used as geometry for the modeling (figure 1**a**). Tumors consist of necrotic ($0 < r/R < 0.4$) and viable regions ($0.4 < r/R < 1$). Tumor tissues are considered as porous media consisting of cancer cells and extracellular space, which is filled with interstitial fluid and extracellular matrix. In the normal tissues, blood vessels and lymphatics are included, though lymphatics are not included in tumor tissues as lymphatics are absent in tumors (2).

### 2.2 Interstitial fluid flow

Interstitial fluid flow is a fluid flow in the tissues which regulates the function of the cells as well as morphogenesis and pathogenesis of tissues (1). In normal tissues, interstitial fluid flows from blood vessels to lymphatic vessels; however, in tumors, lymphatic vessels are lacking (2), and thus, drainage of interstitial fluid to lymphatics is prevented, which leads to elevated interstitial hydrostatic pressure in tumors. Moreover, blood vessels in tumors are leaky (3), which increases infiltration of large molecules from blood vessels to tumor tissues, and thus, increases osmotic colloidal pressure in tumors. Interstitial fluid flow in tumors is mathematically modeled following previous works (4) (5). (6) Interstitial fluid flow in tumor tissues is described by Darcy's law as:

$$u = -K\frac{\partial p}{\partial r} \tag{1}$$

where $u$ [m s$^{-1}$] is flux of interstitial fluid flow, $K$ [m$^2$ Pa$^{-1}$ s$^{-1}$] is hydraulic conductivity of tumor tissues, $p$ [Pa] is interstitial fluid pressure in tumor tissues, and $r$ [mm] is a distance from the periphery. The unit of the flux, [m s$^{-1}$], can be re-written as [m$^3$ m$^{-2}$ s$^{-1}$], to understand its physical meaning of volumetric flow per area. From the continuous equation, the governing equation of interstitial fluid transport is obtained as below:

$$K\frac{\partial^2 p}{\partial r^2} = \underbrace{\Phi_V}_{\substack{\text{Flow from}\\ \text{blood}\\ \text{vessels}}} + \underbrace{\Phi_L}_{\substack{\text{Drainage to}\\ \text{lymphatic}\\ \text{vessels}}} \tag{2}$$

where $\Phi_V$ [s$^{-1}$] is volumetric flux of the interstitial fluid flow from blood vessels to tissues and $\Phi_L$ [s$^{-1}$] is volumetric flux of interstitial fluid flow from tissues to lymphatic vessels. Note $\Phi_L$ is negative as interstitial fluid flows from tissues to lymphatics. Volumetric flux of the fluid source from the blood vessels is described by following Starling's law as:

$$\Phi_V = \frac{J_V}{V} = \frac{L_p S}{V}\underbrace{\left\{\left(p_v - p\right) - \sigma\left(\pi_v - \pi_i\right)\right\}}_{\text{net driving force}} \tag{3}$$

where $J_v$ [m$^3$ s$^{-1}$] is the influx of fluid from blood vessels, $L_p$ [m Pa$^{-1}$ s$^{-1}$] is the hydraulic conductivity of blood vessel walls, $S/V$ [m$^{-1}$] is the surface area of blood vessels per volume, $\sigma$ [-] is osmotic reflection constant, $\pi_v$ and $\pi_i$ [Pa] are the colloidal osmotic pressure of blood vessels and tissues, respectively. Volumetric flux of interstitial fluid to lymphatic vessels as sink term, $\Phi_L$ [m s$^{-1}$], is written in the same manner as:

$$\Phi_L = \frac{J_L}{V} = \frac{L_{pL} S_L}{V} \{ p_L - p - \sigma_L (\pi_L - \pi_i) \} \tag{4}$$

where $L_{pL}$ [m$^2$ Pa$^{-1}$ s$^{-1}$] is hydraulic conductivity of lymphatic vessel walls, $S_L$/V [m$^{-1}$] is the surface area of lymphatic vessels per volume, and $p_L$ [Pa] is interstitial fluid pressure in lymphatic vessels, $\sigma_L$ [Pa] is osmotic reflection coefficient of lymphatic vessels, $\pi_L$ [Pa] is osmotic colloidal pressure in lymphatic vessels. Since lymphatic vessels are absent in tumors (2), $\Phi_L$ is set to zero in tumor tissues. The governing equation of Eqns. (2) – (4) for tumor tissues are re-written as below:

$$\frac{\partial^2 p}{\partial r^2} = \frac{L_p S}{KV} \{ (p_v - p) - \sigma (\pi_v - \pi_i) \} \tag{5}$$

Assuming that parameters $L_p$, $K$, $S/V$, $\pi_v$ , and $\pi_i$ are constant, this equation can be re-written as:

$$\frac{\partial^2 p}{\partial r'^2} = \frac{\alpha^2}{R^2} (p - p_e) \tag{6}$$

where $\frac{\alpha}{R} = \sqrt{\frac{L_p S}{KV}}$ and *r′* is dimensional radius ( $r' = r/R$ ). Note $L_{PL}$ = 0 is assumed. The dimensionless parameter, $\alpha$ , is a measure of the ratio of interstitial to vascular resistances to fluid flow (4). The effective pressure, $p_e$ [mmHg], is the interstitial fluid pressure which would yield zero net volume flux out of vasculature and equal to:

$$p_e = p_v - \sigma(\pi_v - \pi_i) \tag{7}$$

It has been well-known that a decrease in $\alpha$ leads to the decreased interstitial fluid pressure (4). In particular, when $\alpha$ decreases from 50 to 10, interstitial fluid pressure decreases to a small extent. However, when $\alpha$ decreases from 10 to 1, interstitial fluid pressure decreases dramatically (4).

*Effect of collagen content and glycosaminoglycan content on hydraulic conductivity of the interstitium*

The factors that determine hydraulic conductivity of the interstitium are given by Carman-Kozeny equation: the determinants include porosity (defined as fractional void volume, $\phi$), wetted surface area per unit volume (*S*), and a dimensionless proportionality term, the Kozeny factor (G) (9). The role of collagen or glycosaminoglycan (GAG) of extracellular matrix in hydraulic conductivity in tissues has been well-discussed previously (9) (8) (7). Hydraulic conductivity of tissues negatively correlated with GAG content and collagen content (8). The hydraulic conductivities of the interstitium of different tissues with different collagen content and glycosaminoglycan content obtained from previous experimental literature are plotted in figure 2 (9) (8). Hydraulic conductivity in tissues negatively correlated with collagen content and glycosaminoglycan in the interstitium with determination coefficient $R^2$ of 0.86 and 0.80, respectively (figure 2). Thus, the Darcy hydraulic conductivity of the interstitium of the tumors is described with GAG and collagen content as:

$$\log_{10}(\frac{K}{K_0}) = A\log_{10}(\frac{c_c}{c_{c0}}) + B\log_{10}(\frac{c_G}{c_{G0}}) \tag{8}$$

where $c_c$ [mg g$^{-1}$] is collagen content of the interstitium, $c_G$ [mg g$^{-1}$] is glycosaminoglycan content of the interstitium, and $A$ and B are coefficients. This equation is chosen in order to assure the initial hydraulic conductivity at $c_c = c_{c0}$ is the hydraulic conductivity of tumors without any degradation, $K_0$. These coefficients are determined from the literature data (figure 2) using multi-linear regression as $A = -0.865$, $B = -1.16$. Thus, the hydraulic conductivity of tumor tissues at different collagen content and glycosaminoglycan content is calculated using eqn. (8). Note that eqn. (8) cannot be used to predict hydraulic conductivity of the interstitium of different tissues because it is dependent on other factors such as porosity and Kozeny factor as well as collagen and GAG content. In this work, this equation is used to predict the decrease in hydraulic conductivity when collagen content is decreased by bacterial proteolytic activity, as described below.

2.3 Transport of chemotherapeutic agents in tumors

Transport of chemotherapeutic agents in tumors is mathematically modeled by diffusion-advection equation with reactions that includes diffusion, interstitial fluid flow, and extravasation from blood vessels, release from liposomes (Doxil only), and uptake by cancer cells. We followed a previous work for the modeling of drug delivery (4) (6) (10) (11) with modifications from porous media approach, as well as including release term and Doxil and doxorubicin uptake by cancer cells. Cancer cell uptake term is included not just because it is important to evaluate the anti-cancer efficacy of chemotherapeutic agents but it affects diffusion across vascular walls and interstitial fluid flow. We chose *PEG*ylated liposomal doxorubicin (Doxil) and non-liposomal free-doxorubicin as chemotherapeutic agent because the aim of this work is to understand the mechanisms of the enhanced doxorubicin concentration in tumors by the co-administration of Doxil and *C. novyi-NT* (12). Free doxorubicin is also chosen because drug concentration in tumors is not enhanced by the combination of free-doxorubicin and *C. novyi-NT,* in which the mechanisms are unclear. The governing equation of transport of Doxil in tumors is described as:

$$\frac{\partial\left(c^i\phi_l\right)}{\partial t}=\underbrace{\overset{\substack{\text{effective}\\\text{diffusion}\\\text{coefficient}}}{D_{eff}}\frac{\partial^2\left(c^i\phi_l\right)}{\partial r^2}}_{\text{diffusion}}-\underbrace{\frac{\partial}{\partial r}(\overset{\substack{\text{interstitial fluid}\\\text{velocity}}}{u}\;c^i)}_{\text{interstitial fluid flow}}+\underbrace{\overset{\substack{\text{solute flux}\\\text{across vascular walls}}}{J_s}\frac{S}{V}}_{\substack{\text{extravasation from}\\\text{blood vessels}}}-\underbrace{\overbrace{k_{lip}^{uptake}}^{\substack{\text{uptake rate}\\\text{by cancer cells}}}\left(c^i\phi_l\right)}_{\substack{\text{uptake to}\\\text{cancer cells}}}-\underbrace{\overbrace{k_{rel}^{i}}^{\substack{\text{release rate}\\\text{from liposomes}}}\left(c^i\phi_l\right)}_{\text{release from liposomes}} \tag{9}$$

where $c^i$ [mol l$^{-1}$] is Doxil concentration in the interstitium, $D_{eff}$ [m$^2$ s$^{-1}$] is effective diffusion coefficient of liposomal doxorubicin in tumors, $\varphi_l$ [-] is available volume fraction of Doxil, $u$ [m s$^{-1}$] is the velocity of interstitial fluid flow shown in Eq. (1), $J_s$ [mol l$^{-1}$ s$^{-1}$] is the solute flux from blood vessels to tumor tissues, $k_{lip}^{uptake}$ [s$^{-1}$] is uptake rate of liposomes by cancer cells, $k_{rel}^{i}$ [s$^{-1}$] is release rate of doxorubicin from liposomes in the interstitium. Transport of drugs across vasculature from blood vessels, which is described in solute flux of $J_s$ [mol m$^{-2}$ s$^{-1}$] is given by Staverman-Kedem-Katchalsky equation:

$$J_s=\underbrace{J_F(1-\sigma_F)c_v}_{\substack{\text{convection}\\\text{through}\\\text{blood vessel walls}}}+\underbrace{P(c_v-c\;\phi)\frac{Pe}{\exp(Pe)-1}}_{\text{diffusion acorss vascular walls}} \tag{10}$$

where $J_F$ [m s$^{-1}$] is volumetric flux of interstitial fluid through vascular wall (given by eqn. (3))

$J_s=\underbrace{J_F(1-\sigma_F)c_v}_{\substack{\text{convection}\\\text{through}\\\text{blood vessel walls}}}+\underbrace{P(c_v-c\;\phi)\frac{Pe}{\exp(Pe)-1}}_{\text{diffusion acorss vascular walls}}$ and $c_v$ [mol l$^{-1}$] is drug concentration in blood vessels, $P$ [m s$^{-1}$] is vascular permeability of blood vessels for liposomal doxorubicin, $\left(1-\sigma_F\right)$ is the solvent-drag reflection coefficient. The *Peclet* number, the ratio of convection to diffusion, is given as (13):

$$Pe = \frac{L_p\left\{p_v - p - \sigma(\pi_v - \pi_i)\right\}(1-\sigma_f)}{P} \tag{11}$$

On the other hand, concentration of Doxil in cancer cells, $c^c$ [µg ml$^{-1}$], is described as following:

$$\frac{\partial\left(c^{cell}\phi_{cell}\right)}{\partial t} = \underbrace{k_{lip}^{uptake}\left(c^i\phi_l\right)}_{\substack{\text{uptake}\\ \text{from intrestitium}}} \underbrace{-k_{rel}^{c}\left(c^{cell}\phi_{cell}\right)}_{\substack{\text{doxorubicin release}\\ \text{from liposomes}}} \tag{11}$$

where $k_{rel}^{cell}$ [s$^{-1}$] is release rate of doxorubicin from liposomes in cancer cells, $\phi_{cell}$ [-] is volume fraction of cancer cell. The doxorubicin concentration in interstitium is expressed in:

$$\frac{\partial\left(c_f^i\phi_f\right)}{\partial t} = \underbrace{k_{rel}^{i}\left(c_l^i\phi_f\right)}_{\substack{\text{doxorubicin release}\\ \text{from liposomes}}} \underbrace{-k_3\left(k_1c_f^i + k_2\frac{c_f^i}{K_i + c_f^i}\right)\phi_f}_{\text{doxorubicin uptake in cancer cells}} \tag{12}$$

The doxorubicin concentration in cancer cells is also described as following:

$$\frac{\partial\left(c_f^{cell}\phi_c\right)}{\partial t} = \underbrace{k_{rel}^{cell}\left(c^{cell}\phi_c\right)}_{\substack{\text{release}\\ \text{from liposomes}}} + \underbrace{k_3\left(k_1c_f^i + k_2\frac{c_f^i}{K_i + c_f^i}\right)\phi_{cell}}_{\text{doxorubicin uptake in cancer cells}} \tag{13}$$

For the doxorubicin uptake kinetics by cancer cells, we followed a previous work (14). $K_i$ [µg ml$^{-1}$] is the inhibitory constant, $k_1$ [ng $10^5$ cell$^{-1}$ (ug ml$^{-1}$)$^{-1}$], $k_2$ [ng $10^5$ cell$^{-1}$] and $k_3$ [h$^{-1}$] are constant for doxorubicin uptake kinetics by cancer cells. The parameter $k_1$ gives the ratio of intracellular to extracellular concentration at which the net rate of passive exchange is zero. The inhibitory constant is the concentration at which the intracellular uptake saturates. For the details of the kinetic modeling, see (14).

Transport of free-doxorubicin in tumors is also described in the same manner, but it includes the term of doxorubicin uptake by cancer cells instead of liposome uptake and doxorubicin release from liposomes and described as:

$$\frac{\partial\left(c_f^i\phi_f\right)}{\partial t}=\underbrace{D_{eff}^f\frac{\partial^2 c_f^i}{\partial r^2}}_{\text{diffusion}}\underbrace{-\frac{\partial}{\partial r}(uc_f^i)}_{\text{interstitial fluid flow}}+\underbrace{J_{sf}\frac{S}{V}}_{\substack{\text{extravasation from}\\ \text{blood vessels}}}\underbrace{-k_3\left(k_1c_f^i+k_2\frac{c_f^i}{K_i+c_f^i}-c_f^{cell}\right)\phi_{cell}}_{\text{doxorubicin uptake in cancer cells}} \tag{14}$$

where $c_f^i$ [mol l$^{-1}$] is doxorubicin concentration in the interstitium, $c_b$ [mol l$^{-1}$] is doxorubicin concentration binding to cancer cell, $D_{eff}^f$ [m$^2$ s$^{-1}$] is effective diffusion coefficient of doxorubicin in tumors, $\varphi_f$ [-] is available volume fraction of doxorubicin.

The doxorubicin concentration in cancer cells is also described as following:

$$\frac{\partial\left(c_f^{cell}\phi_c\right)}{\partial t}=\underbrace{k_3\left(k_1c_f^i+k_2\frac{c_f^i}{K_i+c_f^i}\right)\phi_{cell}}_{\text{doxorubicin uptake in cancer cells}} \tag{15}$$

The solvent-drag reflection coefficient, $\left(1-\sigma_F\right)$, indicates the ratio of the solute flux across vasculature to the fluid flux.

For the transport of 5-fluorouracil in tumors, the model is based on the free-doxorubicin transport shown in eqn. (14) with modifications of terms of uptake.

$$\frac{\partial\left(c_{fu}^i\phi_f\right)}{\partial t}=\underbrace{D_{eff}^f\frac{\partial^2 c_{fu}^i}{\partial r^2}}_{\text{diffusion}}\underbrace{-\frac{\partial}{\partial r}(uc_{fu}^i)}_{\text{interstitial fluid flow}}+\underbrace{J_{sf}\frac{S}{V}}_{\substack{\text{extravasation from}\\ \text{blood vessels}}}\underbrace{-k_{21}c_{fu}^i\phi_f}_{\substack{\text{binding}\\ \text{on cell membrane}}}\underbrace{+k_{12}c_{fu}^{cell}\phi_f}_{\substack{\text{dissociation from}\\ \text{cell membrane}}} \tag{16}$$

The 5-FU concentration in cancer cells is also described accordingly:

$$\frac{\partial\left(c_{fu}^{cell}\phi_{cell}\right)}{\partial t}=\underbrace{k_{21}c_{fu}^i\phi_{cell}}_{\substack{\text{binding}\\ \text{on cell membrane}}}\underbrace{-k_{12}c_{fu}^{cell}\phi_{cell}}_{\substack{\text{dissociation from}\\ \text{cell membrane}}} \tag{17}$$

where $k_{21}$ and $k_{12}$ [$s^{-1}$] is binding and dissociation constant of 5-fluorouracil on/off cancer cell membrane.

*Effect of geometric tortuosity and available volume fraction on effective diffusion coefficient*

Diffusion coefficient of liposomal doxorubicin in tumors is related to the geometric tortuosity and available volume fraction (porosity), and described as:

$$D_{eff} = \frac{\varphi}{\tau} D_{int} \quad (18)$$

where $\tau$ [-] is tortuosity of tumors, $D_{int}$ [$m^2\ s^{-1}$] is diffusion coefficient of liposomal doxorubicin in the interstitium. The tortuosity value of solid tumors of $\sqrt{2}$ was used, which is the theoretical value for well-packed porous media (20).

*Effect of collagen content on diffusion coefficient of chemotherapeutic agents*

Dense collagen of extracellular matrix is a major barrier to drug penetration in tumors. For example, diffusivity of *IgG* antibody in tumors has been reported to be inversely proportional to collagen content of tumors (21). Thus, the contribution of the effect of collagen content to effective diffusivity of chemotherapeutic agents is included in the modeling. Contribution of tortuosity to diffusivity of particles in tumors is divided to two factors as shown in eqn. (10): geometric factor (geometric resistance due to solid tissues), which was described in eqn. (15), and viscous factor (viscous resistance due to ECM of collagen, proteoglycan), which is to be discussed here.

$$D_{eff}(c_c) = \underbrace{\left( \frac{D_{eff}(c_c)}{D_{int}(c_c)} \right)}_{\substack{\text{geometric} \\ \text{contribution}}} \underbrace{\left( \frac{D_{int}(c_c)}{D_0} \right)}_{\substack{\text{viscous} \\ \text{contribution}}} D_0 \tag{19}$$

The viscous contribution is described using viscous tortuosity parameter $\tau_v(c_c)$ (22) as

$$\frac{D_{int}(c_c)}{D_0} = \frac{1}{\tau_v^{\,2}(c_c)} \tag{20}$$

Here we denote viscous parameter $\tau_v(c_c)$ as a function of collagen content, $c_c$. From eqn. (18) and (20), eqn. (19) can be re-written as follows:

$$D_{eff}(c_c) = \underbrace{\frac{\varphi}{\tau}}_{\substack{\text{geometric} \\ \text{contribution}}} \underbrace{\left( \frac{1}{\tau_v(c_c)^2} \right)}_{\substack{\text{viscous} \\ \text{contribution}}} D_0 \tag{21}$$

The dependence of viscous tortuosity on collagen content was calculated from the diffusion coefficient of particles in collagen gel with different collagen content (20). We here consider the simple diffusion in one-dimensional collagen gel, so the first term of eqn. (21) is assumed to be one in the diffusion in collagen gel. The parameter, $\tau_v(c_c)$, which describes the extent of inhibition of diffusion by collagen, was determined by plotting $\sqrt{D_{eff}/D_0}$ against different collagen content (figure S2). The diameter of Doxil is 80–85 nm, so we used the hydrodynamic radius of 40 nm. Diffusion coefficients under different collagen content were obtained from previous work (20). We calculated the diffusion coefficients with hydrodynamic radius of 40 nm under different collagen contents from the data (20), and viscous tortuosity, $\tau_v(c_c)$, is calculated from figure S2 as,

$$\left(\frac{D_0}{D}\right)^{0.5} = \tau_v(c_c) = \kappa c_c + \gamma \quad (22)$$

where $\alpha$ [ml mg$^{-1}$] and $\beta$ [-] are coefficients for viscous tortuosity. These coefficients are determined from the relationship between effective diffusion coefficient in collagen gels with different collagen content from previous literature (20) figure S2. For the 40 nm particles, which are used for the modeling of Doxil (diameter 85 nm), $\alpha$ = 0.10 [ml mg$^{-1}$] and $\beta$ = 1.03. Thus, the viscous contribution is calculated as:

$$\frac{1}{\tau_v(c_c)^2} = \frac{1}{\left(\kappa c_c + \gamma\right)^2} \quad (23)$$

Finally, the effective diffusion coefficient in tumors is calculated as:

$$D_{eff}(c_c) = \underbrace{\frac{\varphi}{\tau}}_{\substack{\text{geometric}\\ \text{contribution}}} \underbrace{\frac{1}{\left(\kappa c_c + \gamma\right)^2}}_{\substack{\text{viscous}\\ \text{contribution}}} D_0 \quad (24)$$

To validate the model of eqn. (21), the effective diffusion coefficients of 2M-dextran, 76.2 nm-radius liposomes, and IgG are calculated and the obtained diffusion coefficients are compared against the measured ones *in vivo,* as shown in figure S2. For the particles with hydrodynamic radius of 76 nm, the coefficients $\alpha$ and $\beta$ were determined as $\alpha$ = 0.11 [ml mg$^{-1}$] and $\beta$ = 0.84 (20). Predicted diffusion coefficient in tumors agreed reasonably well with experimentally literature (21) (22) (23). Using these values with $\varphi$ = 0.05 and $\tau$ = 1.4, the effective diffusion coefficient in U87 glioblastoma with collagen content of the interstitium of 45 mg ml$^{-1}$ is calculated as 2.24 x 10$^{-13}$ m$^2$ s$^{-1}$. We used the diffusion coefficient of liposomal doxorubicin with diameter 85 nm (2.1 x 10$^{-10}$ m$^2$ s$^{-1}$) for liposomes with 76 nm radius, which we used here for validation. The effective diffusivity in U87 tumor predicted in this model is 2.24 x 10$^{-13}$ m$^2$ s$^{-1}$, which agreed reasonably well with measured diffusion coefficient of liposome with 76.2 nm radius in U87/Mu89 mice, 2.97 x $10^{-13}$ $m^2$ $s^{-1}$ (19). The predicted and measured diffusion coefficient in tumors is shown in Figure S3. Note that the measured diffusion coefficients of liposome and 2M-dextran for U87/Mu89 tumor and that of IgG for HSTS26T is smaller than predicted ones (Fig. S3). This discrepancy is due to the experimental fact that diffusion is more inhibited by collagen in tumors *in vivo* than in collagen gel even at the same collagen content, especially for larger collagen content (20). The collagen content of the interstitium of tumors are obtained from previous literature (21).

2.4 Transport of *Clostridium* in tumors

Transport of *Clostridium* in tumors is mathematically modeled in diffusion-advection equation with appropriate reactions. Motility (random motion of bacteria that repeat run and tumble), interstitial fluid flow, and extravasation from blood vessels, and bacterial growth term are included. Chemotactic term is not included in this modeling for simplicity Motility is expressed in diffusion equation using effective random motility coefficient, $\mu_{eff}$ [$m^2$ $s^{-1}$], and interstitial flow is expressed in convection. Growth term is expressed in Baranyi model (24). Thus, the governing equation of bacterial transport in tumors is described as:

(25)

$$\frac{\partial\left(b\varphi_b^t\right)}{\partial t} = \underbrace{\mu_{eff}\frac{\partial^2\left(b\varphi_b^t\right)}{\partial r^2}}_{\text{Motility}} \underbrace{-\frac{\partial}{\partial r}\{ub\}}_{\text{Interstitial flow}} + \underbrace{J_F(1-\sigma_F)b}_{\substack{\text{convection}\\ \text{through}\\ \text{blood vessel walls}}} + \underbrace{P(b_v - b\,\phi)\frac{Pe}{\exp(Pe)-1}}_{\text{transvascular diffusion}} + \underbrace{\frac{q(t)}{1+q(t)}u_{\max}\left(1-\left(\frac{c}{c_{\max}}\right)\right)b\phi}_{\text{Growth}}$$

where *b* [CFU $ml^{-1}$] is bacterial concentration in tumor interstitium, $\varphi_b^t$ is available volume fraction of *Clostridium* in tumors. Tumor is considered as porous media that consist of interstitial space and cancer cells. Bacterial effective random motility coefficient in tumors, $\mu_{eff}$, is described using porosity for bacteria (available volume fraction), $\varphi_b^t$, and geometric tortuosity, $\tau$, and random motility coefficient in water, $\mu_0$ [m$^2$ s$^{-1}$], as (25) (26):

$$\mu_{eff} = \frac{\varphi_b^t}{\tau}\mu_0 \tag{24}$$

Bacterial growth is expressed as (24):

$$\frac{\partial b}{\partial t} = \underbrace{\alpha(t)}_{\substack{\text{adjustmen}\\ \text{function}}} \ \underbrace{u(b)}_{\substack{\text{inhibition}\\ \text{function}}} b \tag{25}$$

Here the adjustment function, *α(t),* depends on *q(t)*, which represents the physiological state of the cells (24):

$$\alpha(t) = \frac{q(t)}{1+q(t)} \tag{26}$$

The physiological state can be expressed as

$$\frac{\partial q}{\partial t} = \begin{cases} u_{\max} q(t) & (c \geq 10 \text{ CFU ml}^{-1}) \\ 0 & (c < 10 \text{ CFU ml}^{-1}) \end{cases} \tag{27}$$

The adjustment function, as given in Eqn. 26, can be considered a transformation of quantity $q(t)$ and expressed as the same 'readiness' of the cells for the actual environment.

The $u(c)$ function is called the 'inhibition' function because it ensures the transition of the growth curve to the stationary phase:

$$u = u_{\max}\left(1 - \left(\frac{b}{b_{\max}}\right)^m\right) \tag{28}$$

A value of $m$ = 1 was used here, corresponding to the logistic, or Pearl–Verhurst, growth model. The initial value of a physiological state can be calculated as a product of the lag parameter and maximum specific growth rate.

$$q(0) = \frac{1}{e^{h_0} - 1} \tag{29}$$

$$h_0 = \lambda u_{\max} \tag{30}$$

2.5 Pharmacokinetics of Doxil, free-doxorubicin, and *C. novyi-NT* in plasma

Concentration of Doxil, free-doxorubicin, and *C. novyi-NT* in blood vessel plasma is expressed in pharmacokinetics of two-compartment model (27). In two-compartment model, the drug concentration in plasma in blood vessels, $c_v$ [μg ml$^{-1}$], and in tumors (periphery), $c_p$ [μg ml$^{-1}$], are described as follows (27):

$$V_v \frac{dc_v}{dt} = \underset{\substack{\text{Transport}\\\text{from tumors}\\\text{back to}\\\text{blood vessels}}}{V_p k_d c} \; \underbrace{-V_v k_p c_v}_{\substack{\text{Transport from}\\\text{blood vessels}\\\text{to tumours}}} \; \underbrace{-V_v k_{el} c_v}_{\substack{\text{Elimination from}\\\text{blood vessels}}} \tag{31}$$

$$V_p \frac{dc_p}{dt} = \underset{\substack{\text{Transport from}\\\text{blood vessels}\\\text{to tumors}}}{V_b k_p c} \; \underbrace{-V_p k_d c_p}_{\substack{\text{Transport from}\\\text{tumors to}\\\text{blood vessels}}} \tag{32}$$

where $c_v$ [μg ml$^{-1}$] is doxorubicin concentration in blood vessel plasma, $V_v$ [m$^3$] is the volume of blood vessels (1.46 ml in mice), $V_p$ [m$^3$] is the volume of the periphery (i.e. tumors), $k_p$ [h$^{-1}$] is the rate constant for transport from the blood into the peripheral tissues, $k_d$ [h$^{-1}$] is the rate constant for transport from the peripheral tissue back into circulation, and $k_{eli}$ [h$^{-1}$] is elimination constant from blood vessels. These equations are re-formed and represented in second-order differentiation:

$$\frac{d^2 c_v}{dt^2} + A\frac{dc_v}{dt} + Bc_v = 0 \tag{33}$$

The solution for this equation is given as:

$$c_v = Ae^{-\alpha t} + Be^{-\beta t} \tag{34}$$

The first term of eqn. (22) is distribution phase and the second term is elimination phase. Model parameters, *A, B* [μg ml$^{-1}$], $\alpha$, and $\beta$ [h$^{-1}$] are dependent on drug or bacteria. Note the parameters A + B [μg ml$^{-1}$] indicates the injected dose, and $\ln(2)/\alpha$ or $\ln(2)/\beta$ [h$^{-1}$] indicates the half-time for distribution or elimination phase, respectively.

2.6 Bacterial proteolysis kinetics on collagen

Proteolysis on extracellular matrix of collagen by *C. novyi-NT* was included in the modeling and proteolysis reaction rate is expressed in zero-order kinetics as following:

$$\frac{\partial c_c}{\partial t} = \frac{c_c V_{max}}{K_m + c_c} \tag{35}$$

where $V_{max}$ [mg ml$^{-1}$ min$^{-1}$] is maximum reaction rate and $K_m$ [mg ml$^{-1}$] is Michaelis constant. *C. novyi-NT* secrete collagenase *ColG,* class I, M9B family, which degrades collagen type I, II, and III (28). *C. novyi-NT* do not secrete hyaluronidase (29) because the *C. novyi* strains that secrete hyaluronidase are all type B strains (30), while *C. novyi-NT* is a type A strain.

2.5 The effect of solid stress alleviation by bacterial proteolysis on collagen in drug delivery

Solid stress, which is a physical force that generates in the tumors by dense extracellular matrix of both collagen and hyaluronan, is known to be involved in cancer progression and therapeutic efficacy by compressing blood vessel in tumors (11). Compressed blood vessels decrease perfusion and/or vascular density, and thus, reduces drug delivery. Decompressing blood vessels by depleting collagen and/or hyaluronan can improve drug delivery because it improves blood vessel perfusion or vascular density (11). We included the role of solid stress alleviation in the mathematical modeling because extracellular matrix of collagen is degraded by bacterial proteolytic activity, which alleviates solid stress of tumors. For simplicity, in our modeling, the effect of solid stress alleviation on drug delivery is included by changing blood vessel density in necrotic areas (vessel area per volume, S/V) because of the following reasons. Previous modeling work showed that when solid stress reached a threshold value, the blood vessels collapse and vascular density decreases (66). In our work, since all degradable collagen of type I, II, III is degraded by bacterial proteolysis before drug injection, the surface area of blood vessel per volume is changed as following:

$$\frac{S}{V} = 2000 \ (r/R < 0.4) \ \text{(without bacteria)} \qquad (36)$$

$$\frac{S}{V} = 12000 \ (r/R < 0.4) \ \text{(with bacteria)} \qquad (37)$$

The selection of vessel density value is described in Sec. 2.7.

2.6 Initial and boundary conditions:

2.6.1 Interstitial fluid flow

Boundary conditions for tumor surrounded by normal tissues:

Since the flux and pressure are equal at the boundary of tumor and normal tissues, the boundary conditions are as follows:

$$-K_T \frac{dp}{dr} = -K_N \frac{dp}{dr} \qquad (38)$$

$$p \ (r = R\text{-}) = p \ (r = R+) \qquad (39)$$

where $K_T$ is the hydraulic conductivity at the periphery of tumors and $K_N$ is the hydraulic conductivity at the periphery of normal tissues. R- and R+ indicate the radius of *R* of tumors and that at normal tissue, respectively. The boundary condition for the transport of Doxil or doxorubicin, and *Clostridium* is no flux at *r* = 0, and *r* = 2*R*. Also, at the boundary between tumor and normal tissues at *r* = *R*,

$$\left( \underbrace{-D_{eff} \frac{\partial c_f^i}{\partial r}}_{\text{diffusion}} \underbrace{-K \frac{\partial p}{\partial r} c^i}_{\text{interstitial fluid flow}} \right)(r = R-) = \left( \underbrace{-D_{eff} \frac{\partial c_f^i}{\partial r}}_{\text{diffusion}} \underbrace{-K \frac{\partial p}{\partial r} c^i}_{\text{interstitial fluid flow}} \right)(r = R+) \quad (40)$$

The collagen and hyaluronan content of tumor interstitium is at 9.0 mg $ml^{-1}$ and 2.0 mg $ml^{-1}$, respectively, as initial condition (21). The governing equations above are solved by finite-element method using COMSOL Multiphysics 5.0. The simulation results of concentration of Doxil or doxorubicin in tumors are validated against experimental literature data (12). The simulated bacterial concentration in tumors is validated against experimental literature (31) (42).

## 2.7 Parameters chosen in this work

Parameters used in this work are summarized in table 2. Most parameters used in this work are those of colorectal tumor because the simulation results are validated against experimentally measured drug concentration in colorectal tumor *in vivo* (12).

The hydraulic conductivity of normal tissues of 4.2 x $10^{-10}$ $m^2$ $Pa^{-1}$ $s^{-1}$ is used (32), which is larger than tumors at 2.38 x $10^{-11}$ $m^2$ $mmHg^{-1}$ $s^{-1}$ (33) because of the following reasons: (i) tumor tissues are less porous than normal ones due to a rapid proliferation of cancer cells, which leads to reduced hydraulic conductivity, (ii) hydraulic conductivity is negatively correlated with glycosaminoglycan, as discussed above, and tumor tissue is usually rich in collagen content. Vascular permeability of Doxil at 4.0 x $10^{-10}$ m $s^{-1}$ is used because vascular permeability of liposomes with diameter 90 nm is six times smaller than BSA in human tumor xenograft (34) and vascular permeability of BSA of LS174T in mice is at 2.4 x $10^{-9}$ m $s^{-1}$ (35). Available volume fraction for *Clostridium* in tumors of 0.05 is used because of the following reasons. Available volume fraction for 1 μm particles is not available from literature. *Clostridium* are rod-shaped and their size is 0.5 – 1.9 μm with length 3.0 – 16.9 μm. Available volume fraction for 100 nm radius particles is 0.05 (19). For the tortuosity, $\tau = 1.4$ was used as described in the previous section. Random motility coefficient of *Clostridium* is unavailable from literature. Random motility coefficient of *Salmonella typhimurium* in water at 25°C is reported to be 6.0 x $10^{-9}$ $m^2$ $s^{-1}$ (36). Since random motility coefficient at 37°C is not available, we estimated it as following. Random motility coefficient of bacteria is described using individual cell properties such as swimming speed, *v* [μm $s^{-1}$], tumbling frequency, $p_0$ [$s^{-1}$] and directional persistence, $\phi$ [-] (37), as:

$$\mu_0 = \frac{v^2}{p_0(1-\phi_d)} \tag{41}$$

The index of directional persistence, $\phi_d$ , accounts for the angle a cell's path takes between adjacent runs. In three dimensions, $\phi_d$ is equivalent to the mean of the cosine of the run-to-run angle. If the angle between runs is random, the mean run-to-run angle will be 90 degrees and $\phi_d$ will be equal to zero (38). Thus, we assumed the random angles for the tumbling, and thus, $\phi_d = 0$ is assumed here. Random motility coefficient at 37°C is calculated using that at 25°C from swimming speed and tumbling frequency as:

$$\mu_{37} = \frac{v_{37}^{2}}{v_{25}^{2}} \frac{p_{25}}{p_{37}} \mu_{25} \tag{42}$$

The numbers denoted with subscript indicate the temperature in Celsius. Using the swimming speeds and tumbling frequencies at 25 and 37°C (39), random motility coefficient at 37°C is calculated at 5.1 x $10^{-9}$ $m^2$ $s^{-1}$, which used for the modeling. The vascular permeability of *Clostridium* is not available from literature. *Clostridium* is rod-shaped with approximately 0.5 $\mu$ m x1.5 µm. However, the pore cut-off size of LS174T colorectal tumor blood vessels, which describes the functional upper limit of the size of a particle that can extravasate from the micro-vessels, is 400 – 600 nm (40) and no 800-nm microspheres were seen to extravasate from vessels in non-superfused tumors (412). Thus, we set the vascular permeability of *Clostridium novyi-NT* at zero for simplicity (that is, no passive transport across vascular walls). The growth parameters for Baranyi model are determined from previous experimental literature (42). The maximum growth rate, $\mu_{max}$ [$h^{-1}$], is determined from the logarithm of bacterial concentration when it is linear against time before inhibition period. Lag period is assumed to be zero because the measured bacterial concentration in previous literature incudes the contribution of extravasation from vessels, thus, lag period cannot be determined. This is discussed in results and discussions part in main article.

*Pharmacokinetics*

For pharmacokinetic of liposomal doxorubicin and free doxorubicin, model parameters were obtained from previous experimental work (45). The pharmacokinetic parameters were calculated from experimental data of literature to fit them with two-compartment model, following a previous work for curve fitting (43). Pharmacokinetic parameters of *C. novyi-NT* in plasma were obtained from the measured concentrations of *C. novyi-NT* spores in plasma in CT26 tumor-bearing BALB/c mice (31) and calculated as A = 2.83 x $10^8$ CFU $ml^{-1}$, B = 1.72 x $10^7$ CFU $ml^{-1}$ $\alpha$ = 0.141 $h^{-1}$ and $\beta$ = 0.0019 $h^{-1}$.

*The ratio of collagen content of type I, II, and III to total interstitial collagen*

Extracellular matrix of collagen consists of type I, II, III, and V collagen. Type IV collagen composes basement membrane of blood vessel. *C. novyi-NT* secrete collagenase *ColG,* class I, M9B family, which degrades collagen type I, II, and III (28) (47). The total interstitium collagen content is 9.0 mg $ml^{-1}$ (21); however, the ratio of collagen type I, II, and III among total interstitial collagen (I, II, III, and V) in colorectal tumor is missing. In normal colon, type I collagen occupies 62%, type III 20%, and type V 12% (44). In tumors, the increase in collagen type I synthesis has been reported (48), which altered type I: type III collagen 5:1 instead of 3:1 in colon fibroblast (46). Additionally, collagen type II ratio over type I, III, and Vin lung cancer is 26% (49). Thus, the collagen type ratio of interstitium in colorectal tumors is calculated as type I 61.9%, type II 20.6%, type III, 10.9%, and type V 6.5%. Finally, using the total collagen content of 9.0 mg $ml^{-1}$ for LS174T colorectal tumors (21), the collagen content in the interstitium that can be degraded by *C. novyi-NT*, that is, type I, II, and III collagen, is 8.42 mg $ml^{-1}$. The collagen type V ratio of 6.5% calculated here is close to previous literature showing that the ratio of type V to the total type I, III, and V collagen in breast cancer is approximately 0.10, though type II collagen is not included, (50).

*Collagenolytic activity of C. novyi-NT*

The collagenolytic reaction rate of *Clostridium* is calculated as following.

Collagenolytic activities of *C. tetani* culture were previously reported as follows . The hydroxyproline amount released after 36 h from the mixture of broth, *Clostridium* culture, and collagen solution (5 ml thioglycolate broth, 1 ml pre-gelled collagen, and 0.2 ml bacterial culture) is measured as 66% (52), which estimates the collagenolytic activity of 14.8 μg $ml^{-1}$ $min^{-1}$. The measured collagenolytic activity of collagenase of 0.015% (30 ug in 0.2 ml) at 88% also justifies this value with in vivo experimental results as follows. The exponential decay time, $\tau$ [min] that was measured from the decrease in second harmonic generation (SHG) signal *in vivo* was reported to be 9.9 min, 1.2 x $10^2$ min, and 3.7 x $10^3$ min, for collagenase concentration of 10, 1.0, and 0% (no collagenase), respectively (51). The maximum reaction rate, $V_{max}$, for these collagen degradations was calculated as:

$$\frac{dc_c}{dt} = -\frac{c_c V_{max}}{K_m + c_c} \tag{38}$$

Here we assume an exponential decay, as observed in previous work (51), and thus, the decay constant, k [$h^{-1}$] is calculated as:

$$k = \frac{V_{max}}{K_m + c_c} = \frac{V_{max}}{c_c} = \frac{1}{\tau} \tag{40}$$

Here Michaelis constant, $K_m$, was assumed to be zero because SHG signal decreased in a constant rate during the measured period, which means $K_m$ is small enough compared with the collagen content. The constant collagen content of the interstitium ($c_c$) is assumed at 45 mg $ml^{-1}$ for Mu89 (21) for simplicity. Using these values, the maximum reaction rate, $V_{max}$, is calculated as 4.53 mg $ml^{-1}$ $min^{-1}$ and 0.363 mg $ml^{-1}$ $min^{-1}$ for 10% and 1.0% collagenase, respectively. Note the decrease in collagen content by 0% collagenase was extracted from each decrease in collagen content because this decrease is considered to be due to background difference. Assuming the linearity of bacterial concentration to maximum reaction rate, the estimated maximum reaction rate in vivo for 0.015% collagenase is 6.79 μg

ml$^{-1}$ min$^{-1}$, which is close to the estimated value above. Considering that 52% of hydroxyproline is released at 36 h from figure in literature (52) makes the value more believable.

*The effect of solid stress alleviation by bacterial proteolysis on collagen on drug delivery*

We increased the blood vessel density of whole tumors by 10% because of the following reasons: (i) *ex vivo* collagenase treatment decreased solid stress of tumors to 43%, 41%, 11%, 29% (median 34.9%), especially at the center of tumors (67), (ii) *Saridegib* increased vessel density by approximately 10%, in which solid stress was reduced to 45% – 39% (11). From above, the vessel density in tumor necrotic regions ($0 < r/R < 0.4$) is changed from 2000 m$^{-1}$ to 12,000 m$^{-1}$ after the degradation of extracellular matrix of collagen, which satisfies the total increase in vessel density by 10%.

Supporting figures

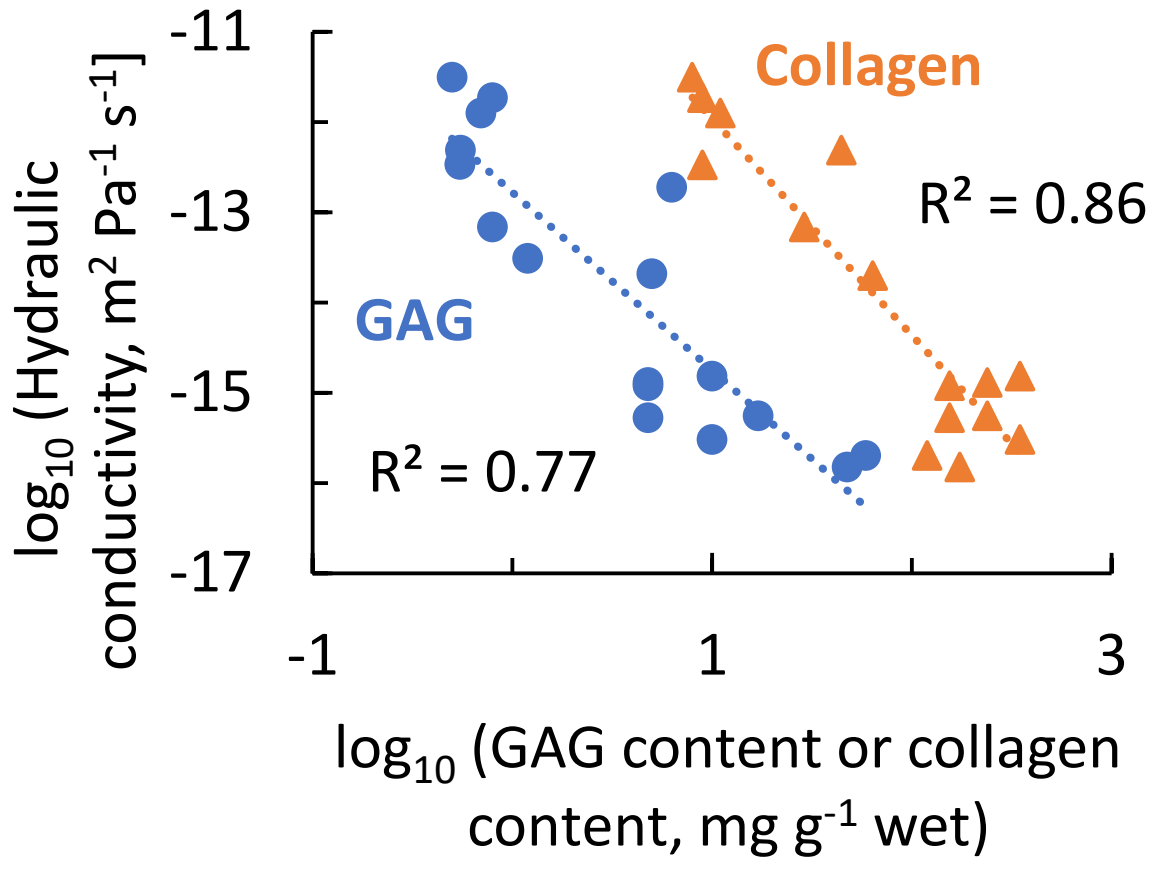

Figure S1. Darcy hydraulic conductivity of tissue are negatively correlated to glycosaminoglycan (GAG) or collagen contents of the interstitium. Data are derived from previous experimental literature (8) (21) (53).

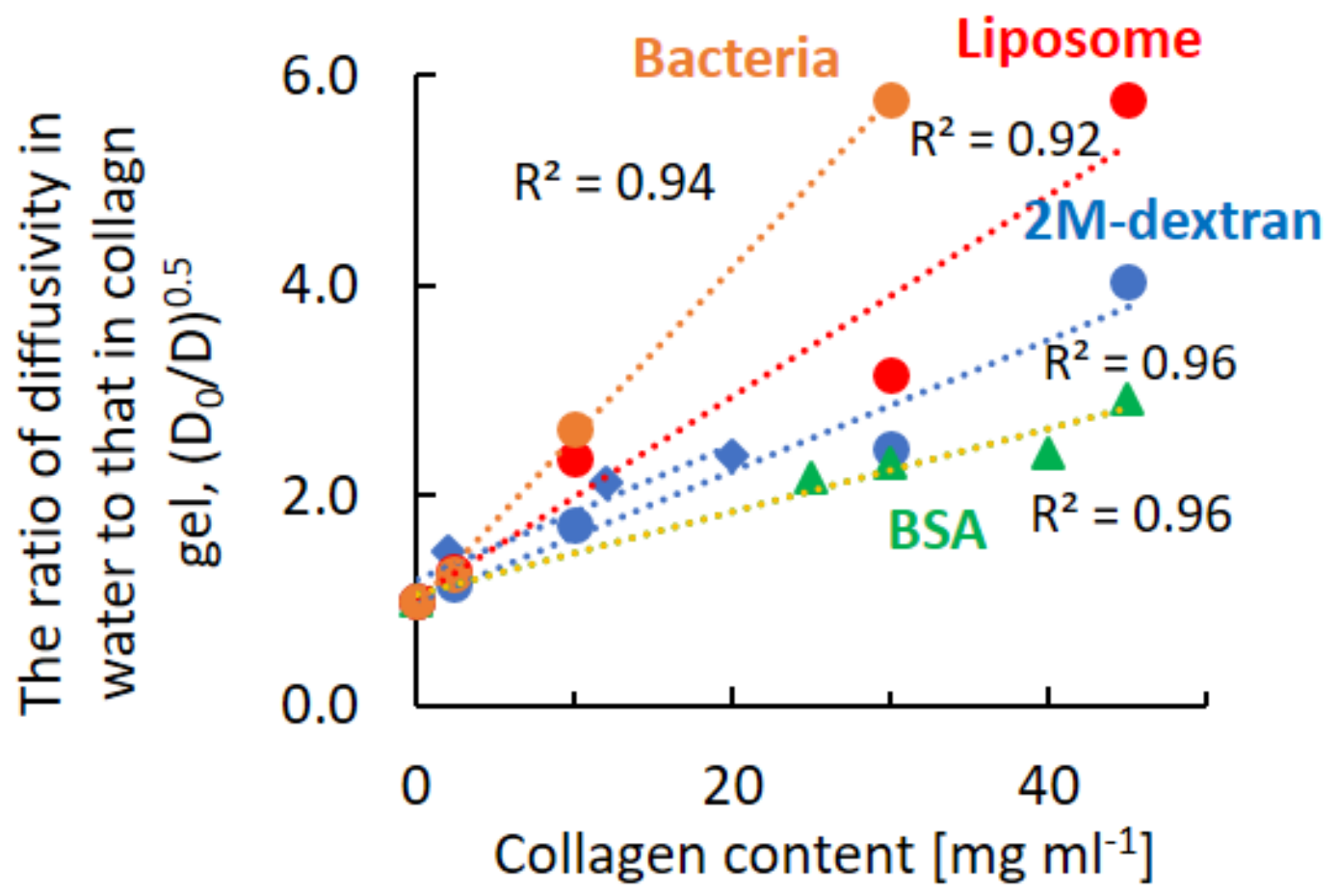

Figure S2. Diffusion of larger particles in collagen gel is more inhibited by collagen than that of smaller ones. $D_0$ [$m^2$ $s^{-1}$]: diffusivity in water; $D$ [$m^2$ $s^{-1}$]: diffusion coefficient in collagen gel at 37°C. Data are calculated from diffusivity of particles in collagen gel with different collagen content. Symbol color indicates the type of particles: diffusivity of bacteria (1 µm sphere) is shown in orange, liposomes with diameter 76 nm is in red, 2M-dextran with radius 22 nm is in blue, and BSA with radius 3.5 nm is green. Data in circle, (20) in diamonds (54), in triangles are obtained from previous literature

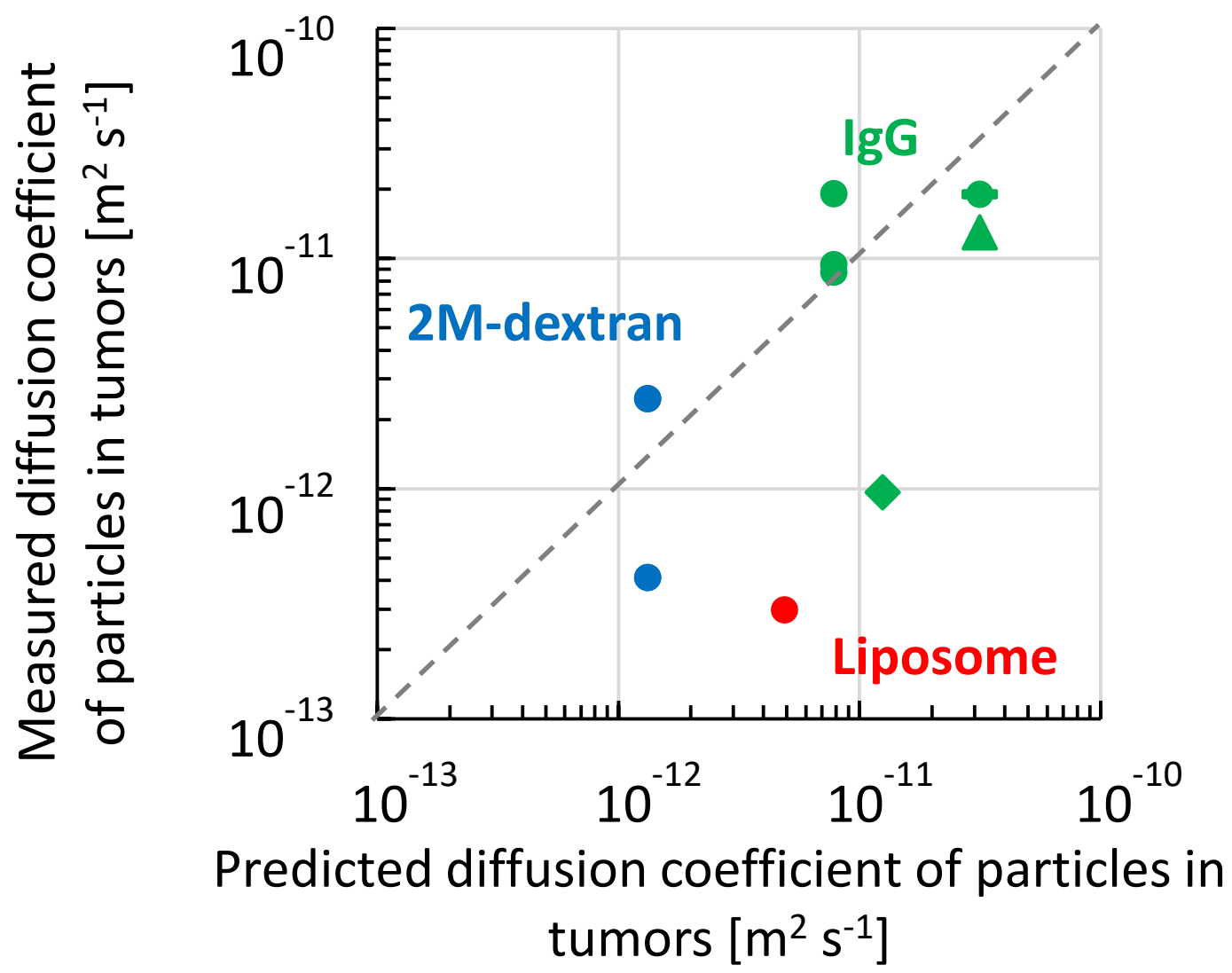

Figure S3. Predicted diffusion coefficient of particles in tumors in this model agreed well with experimental literature (21) (22) (23). The symbol indicates the tumor type: circle is U87/Mu89, diamond is HSTS26T, and triangle is LS174T tumor.

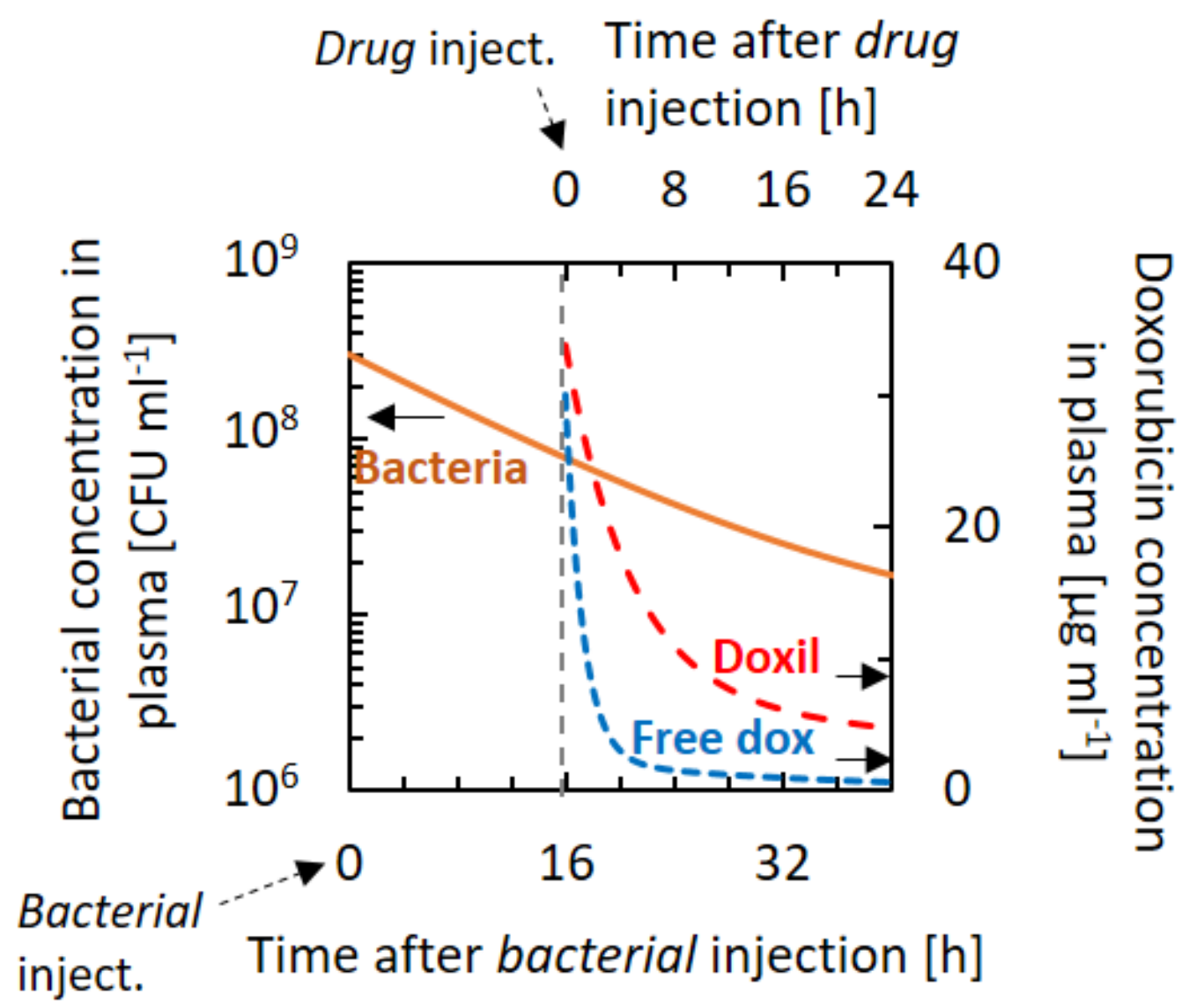

Figure S4. Doxil remains at high concentration in plasma longer than free-doxorubicin (Doxil red dashed-line; free-doxorubicin: Blue dotted-line). The concentration of *C. noyvi-NT* declined rapidly in the plasma (orange solid-line). The bacterial concentration is in logarithmic scale.

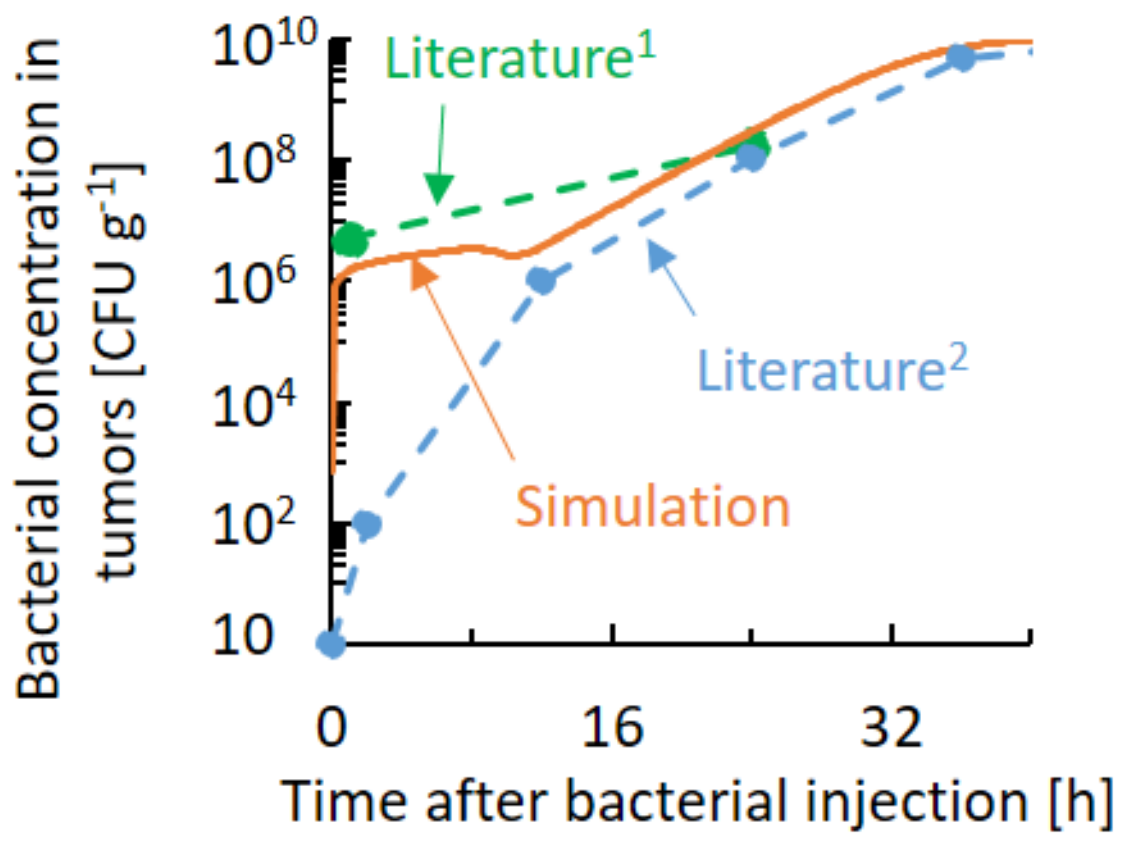

Figure S5. Bacterial concentration in tumors increases rapidly to over 106 CFU ml$^{-1}$ due to extravasation from blood vessels, then it increases gradually. After 12 h, it increases again due to bacterial growth. Shrinkage of figure 7 in vertical axis.

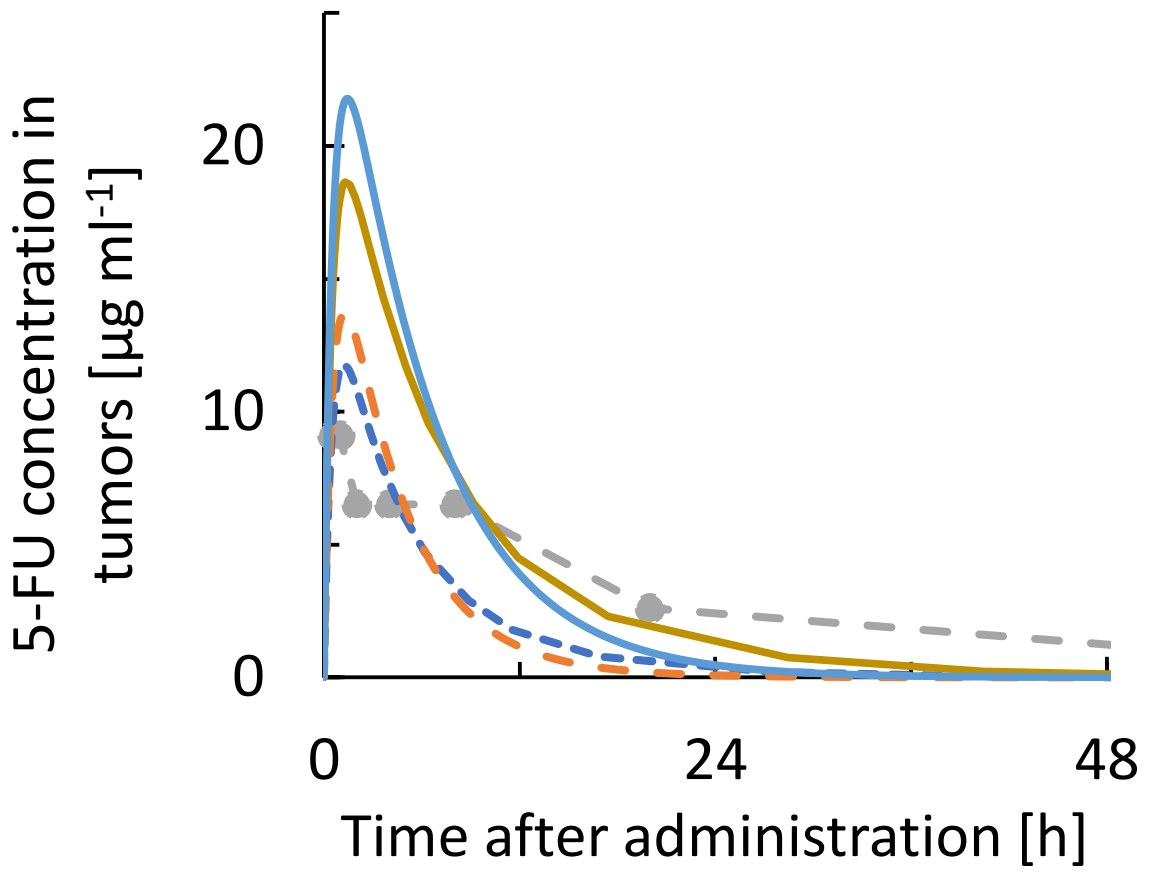

Figure S6. Simulated concentration of 5-fluorouracil in tumors with bacteria (dashed-line in blue) is almost as same as that without bacteria (orange). Gray-dashed line: literature data (41), blue solid-line: x10 binding rate for 5-FU with bacteria, orange solid-line: x10 binding rate for 5-FU alone.

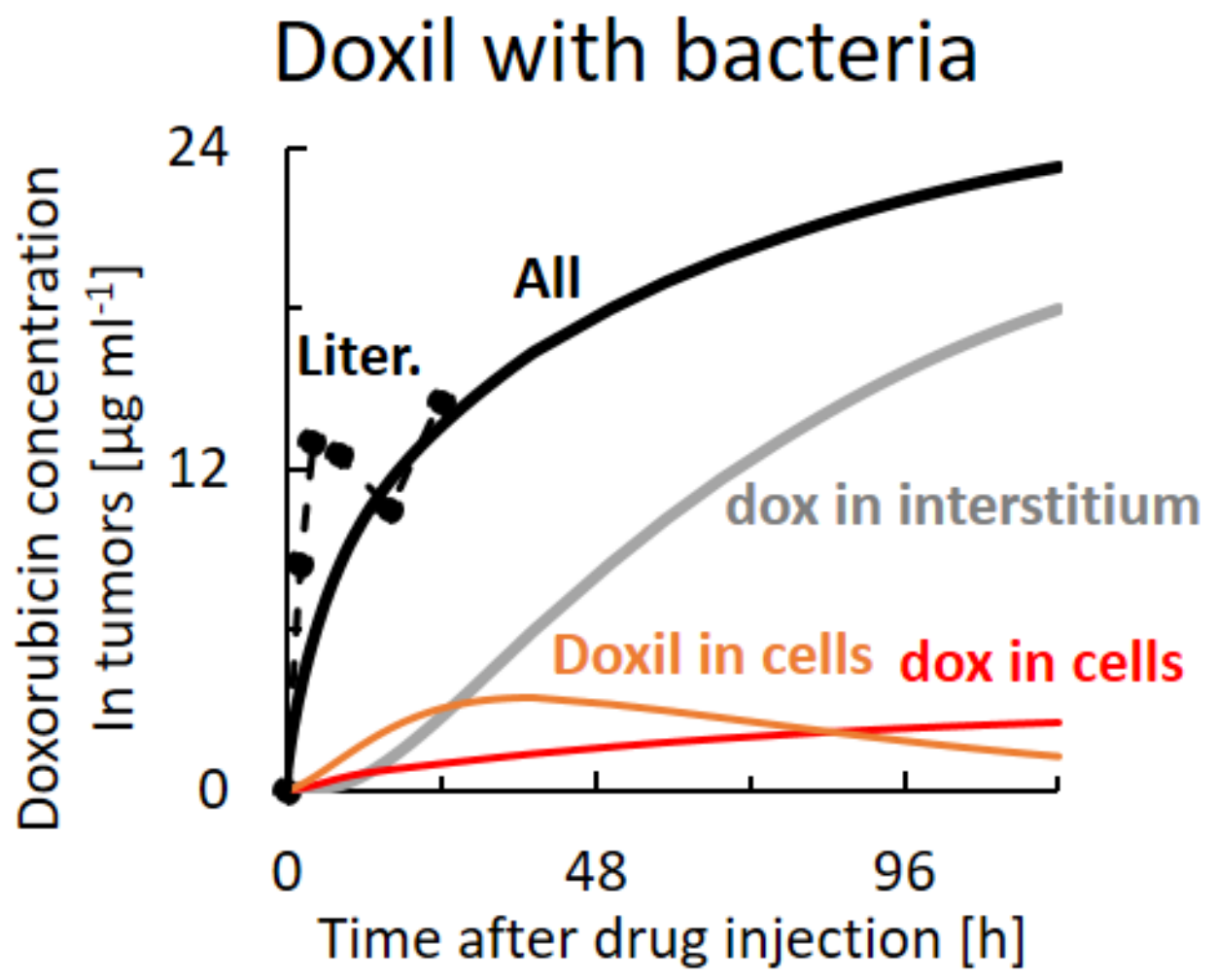

Figure S7. Simulated concentration of all doxorubicin in tumors (light blue), Doxil in cancer cells (orange), doxorubicin in the interstitium (gray), and in cancer cells (red) for Doxil with bacteria. An enlargement in vertical axis of the figure 8**b**.

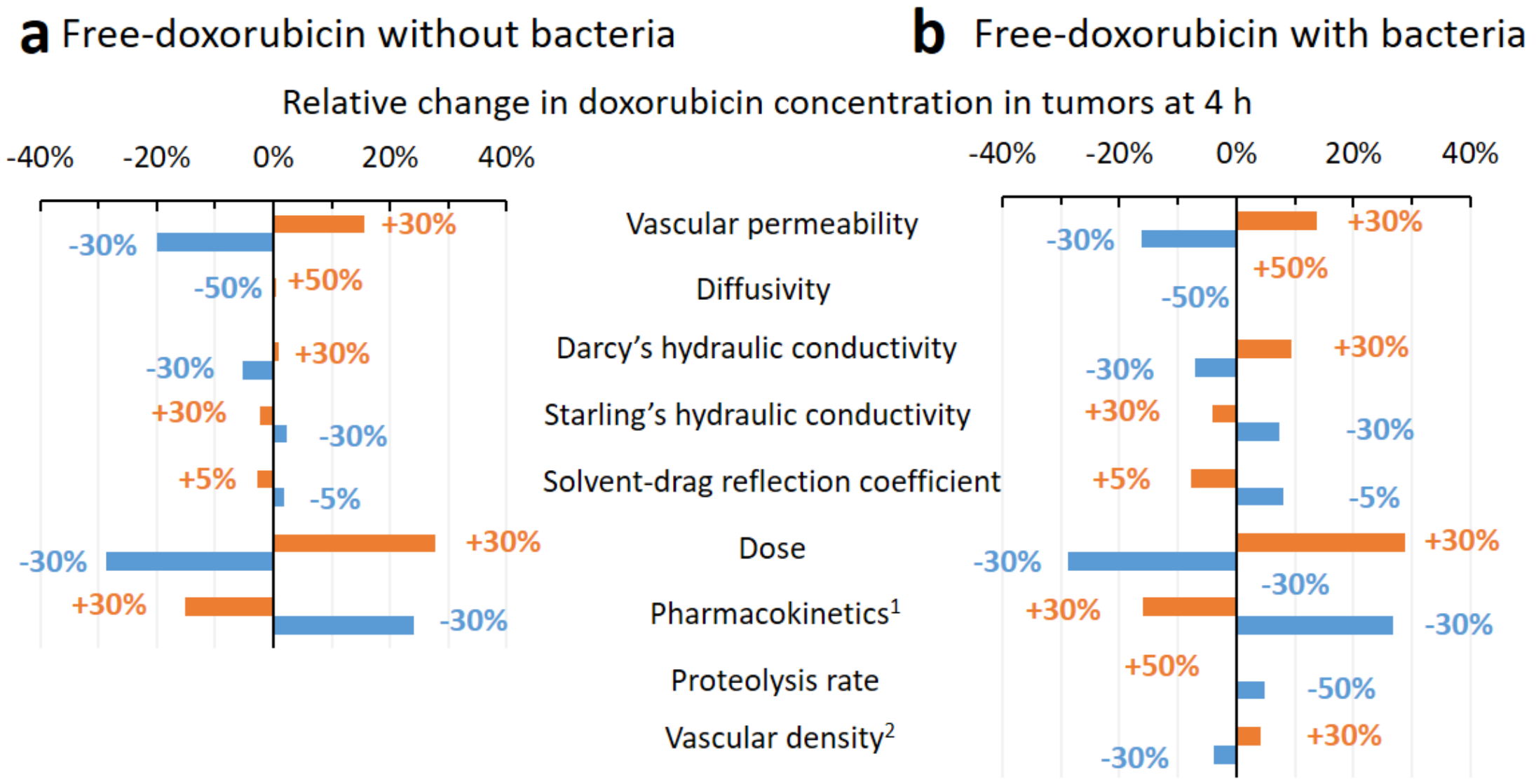

Figure S8. Parametric sensitivity analysis of free-doxorubicin without bacteria (a) and with bacteria (b) in determining doxorubicin concentration in tumors at 4 h.

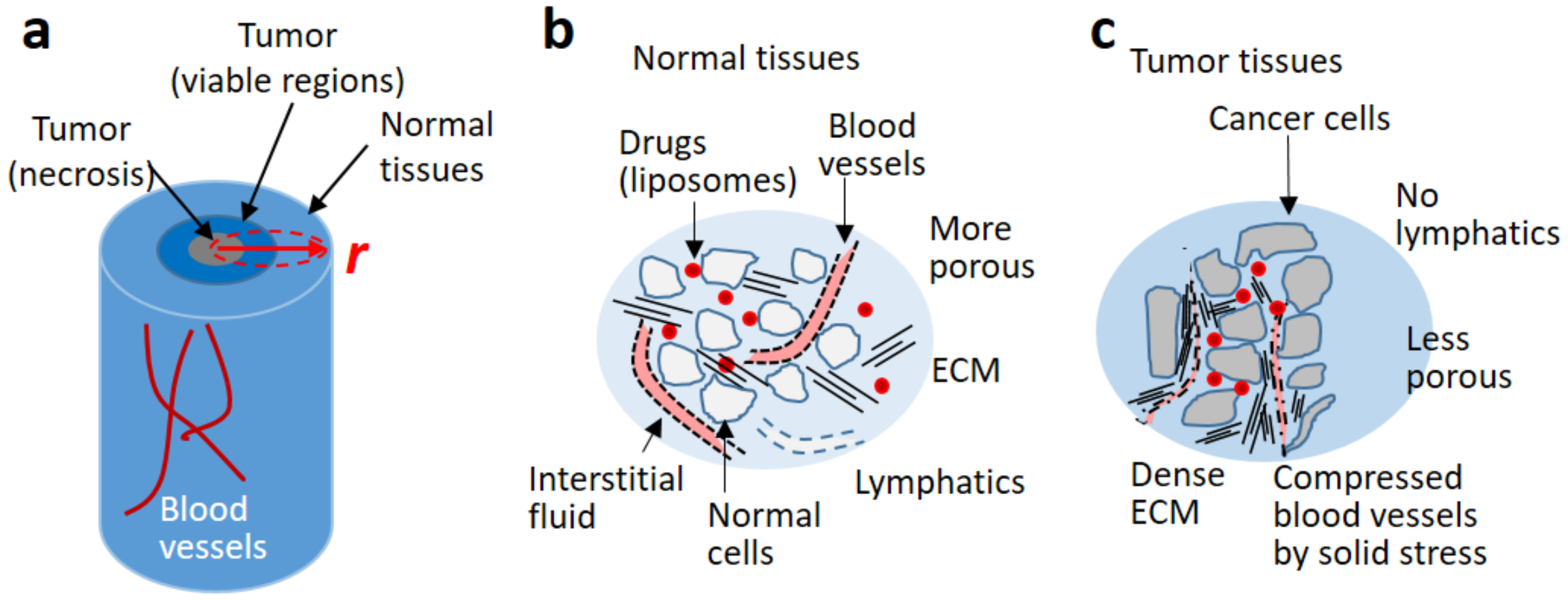

Figure S9. A geometry of tumors and normal tissues (**a**). Tumor consists of necrotic ($0 < r/R < 0.4$) and viable region ($0.4 < r/R < 1$), surrounded by normal tissues ($1 < r/R < 2$). Tissues are considered as porous media consisting of cancer or normal cells and remaining interstitium (**b**, **c**). Normal tissues include functional blood vessels and lymphatics (**b**), while blood vessels in tumors are leaky and compressed by solid stress and lymphatic vessels are absent (**c**).

Table S1. Parameter list

1 mmHg = 101.3 Pa

| Symbols | Parameters | Values | Tumor type, bacterial strain, reference |
|---|---|---|---|
| *A, B* | Y-intercept of pharmacokinetics of Doxil and doxorubicin in plasma | Doxil: A = 27.11 μg $ml^{-1}$, B = 6.69 μg $ml^{-1}$; Free doxorubicin: A = 27.66 μg $ml^{-1}$; B = 2.27 μg $ml^{-1}$ | Dose: 5 mg/kg for both; (45) |
| $A_b$, $B_b$ | Y-intercept of pharmacokinetics of *C. novyi-NT* in plasma | A = 2.83 x $10^8$ CFU $ml^{-1}$, B = 1.66 x $10^7$ CFU $ml^{-1}$ | *C. noyvi-NT*, CT26 tumor-bearing BALB/c mice; Dose: 1.5 x $10^{10}$ CFU/kg; (31) |
| $A_{fu}$, $B_{fu}$ | Y-intercept of pharmacokinetics of 5-fluorouracil in plasma | Doxil: A = 99.5 μg $ml^{-1}$, B = 1.433 μg $ml^{-1}$ | 5-fluorouracil, colorectal tumor-bearing BALB/c mice; Dose: 100 mg/kg.; Peters et al. (1993) |
| *b* | Bacterial concentration in interstitium | Eqn. (25) | |
| $c^{cell}$ | Concentration of Doxil in cancer cells | Eqn. (11) | |
| $c^{i}$ | Concentration of Doxil in interstitium | Eqn. (9) | |
| $c_c$ | Collagen content of interstitium | Eqn. (30) | |
| $c_f^{cell}$ | Concentration of free-doxorubicin in cancer cell | Eqn. (13), (15) | |
| $c_f^{i}$ | Concentration of free-doxorubicin in interstitium | Eqn. (12), (14) | |
| $c_{fu}^{i}$ | Concentration of 5-FU in interstitium | Eqn. (16) | |
| $c_{fu}^{cell}$ | Concentration of 5-FU in cancer cells | Eqn. (17) | |
| $c_{c0}$ | Total collagen content of the tumor interstitium | 9.0 mg $ml^{-1}$ | LS174T (21); |
| $c_{cV}/c_{c0}$ | Type V collagen ratio over total collagen | 6.5% | (55) |
| $c_{h0}$ | Hyaluronan content of tumor interstitium | 0.20 mg $ml^{-1}$ | LS174T; (21) |
| $c_{GAG0}$ | Glycosaminoglycan content of tumors | 0.35 mg $ml^{-1}$ | LS174T; (21) |
| $D_0$ | Diffusion coefficient of Doxil in water at 37°C | 2.1 x $10^{-10}$ $m^2$ $s^{-1}$ | (56) |
| | Diffusion coefficient of free doxorubicin in water at 37°C | 8.83 x $10^{-10}$ $m^2$ $s^{-}$ | (62) |
| $D_{eff}$ | Effective diffusion coefficient | Eqn. (18)–(22) | |
| $K_0$ | Darcy hydraulic conductivity in tumors | 2.38 x $10^{-11}$ $m^2$ $mmHg^{-1}$ $s^{-1}$ | LS174T colorectal cancer; (33) |
| | Darcy hydraulic conductivity in normal tissues | 2.5 x $10^{-11}$ $m^2$ $mmHg^{-1}$ $s^{-1}$ | (58) |
| $k_1$, $k_2$, $k_3$ | Kinetic parameters of doxorubicin uptake rate by cancer cells | $k_1$ = 0.00631 ng $10^5$ $cell^{-1}$ (ug $ml^{-1})^{-1}$; $k_2$ = 0.126 ng $10^5$ $cell^{-1}$; $k^3$ = 1.01 $h^{-1}$ | (14) |
| $k_{21}$, $k_{12}$ | Binding and dissociation constant of 5-fluorouracil on cell membrane | $k_{21}$ = 0.2 $s^{-1}$, $k_{12}$ = 4.1 $s^{-1}$ | (59) |
| $k_{lip}^i$ | Release constant of Doxil from liposomes in the interstitium | 1.77 x $10^{-4}$ $min^{-1}$ | 4T1 breast cancer; (60) |
| $k_{lip}^c$ | Release constant of Doxil from liposomes in cancer cell | 9.6 x $10^{-4}$ $min^{-1}$ | 4T1 breast cancer; (60) |
| $K_m$ | Michaelis constant of proteolytic reaction rate | 0 | See text |
| $k_{uptake_lip}$ | Uptake rate of Doxil by cancer cells | 1.0 x $10^{-5}$ $s^{-1}$ | B16F10 murine melanoma cell; (61) |
| $n_v$ | Number of cancer cells per volume | 6 x $10^8$ cell $ml^{-1}$ | |
| $L_p$ | Hydraulic conductivity of microvascular wall | Tumors: 1.86 x $10^{-8}$ m $mmHg^{-1}$ $s^{-1}$ | (58) |
| | | Normal tissues: 3.6 x $10^{-10}$ m $mmHg^{-1}$ $s^{-1}$ | (58) |
| $P$ | Vascular permeability in tumor vessels | Doxil: 4.0 x $10^{-10}$ m $s^{-1}$; | LS174T colorectal cancer, (34) |
| | | Free doxorubicin: 2.4 x $10^{-9}$ m $s^{-1}$ | LS174T colorectal cancer (Dorsal skinfold chamber); (35) |
| | | *Clostridium novyi-NT*: 0 | 0 (see text) |
| $p_V$ | Capillary pressure | 13.5 mmHg | (63) |
| $q(t)$ | Physiological state of cells | Eqn (27) | |
| $r$ | Distance from tumor core | | |
| $R$ | Radius of tumors | 4 mm | |
| $S/V$ | The ratio of surface area of blood vessels to volumes | Tumor tissue: 20000 $m^{-1}$ (viable regions, $r/R > 0.4$); 2000 $m^{-1}$ (necrotic regions, $r/R < 0.4$ ) | (3) |
| | | Normal tissue: 7000 $m^{-1}$ | |
| $S_L/V$ | The ratio of surface area of lymphatics to volumes in normal tissues | 4300 $m^{-1}$ | (64) |
| $V_{max}$ | Maximum reaction rate | 14.8 ug ml$^{-1}$ | (52) |
| $V_p$ | Volume of tumors | 267.9 mm$^3$ | |
| $V_v$ | Volume of blood vessel | 1.46 ml | |
| $\alpha$, $\beta$ | Pharmacokinetic coefficient of drugs for distribution and elimination phase | Doxil: 0.209 h$^{-1}$; $\beta$ = 0.0169 h$^{-1}$<br>Free doxorubicin: $\alpha$ = 0.778 h$^{-1}$ $\beta$ = 0.0529 h$^{-1}$ | Murine hepatocarcinoma cell (H22) , Dose: 5 mg/kg; (45) |
| $\alpha_b$, $\beta_b$ | Pharmacokinetic coefficient for distribution and elimination phase of *C. noyvi-NT* in plasma | $\alpha_b$ = 0.139 h$^{-1}$; $\beta_b$ = 0.0119 h$^{-1}$ | *C. noyvi-NT*, CT26 tumor-bearing BALB/c mice; Dose: 1.5 x 10$^{10}$ CFU/kg; (31) |
| $\alpha_{fu}$, $\beta_{fu}$ | Pharmacokinetic coefficient for distribution and elimination phase of 5-fluorouracil in plasma | $\alpha_b$ = 1.79 h$^{-1}$; $\beta_b$ = 0.188 h$^{-1}$ | 5-FU, colorectal tumor-bearing BALB/c mice; Dose: 100 mg/kg.; Peters et al. (1993) |
| $\alpha(t)$ | Adjustment function of bacterial growth | Eqn. (26) | |
| $\kappa$ , $\gamma$ | | Doxil: $\kappa$ = 0.10 ml mg$^{-1}$, $\gamma$ = 1.04; | (20) |
| | Coefficient for determining diffusion coefficient at different collagen content | Free doxorubicin: $\kappa$ = 0.04 ml mg$^{-1}$, $\gamma$ = 1.06; | (62) |
| | | Bacteria (1 µm diameter particle): $\kappa$ = 0.16 ml mg$^{-1}$, $\gamma$ = 0.97; | extrapolated from (20) |
| $\mu$ | Random motility coefficient of *Salmonella typhimurium* at 37°C | 5.1 x 10$^{-9}$ m$^2$ s$^{-1}$ | (353) (36) (39) |
| $\sigma$ | Osmotic reflection coefficient | Tumors: 0.91; Normal tissues: 8.7x 10$^{-5}$ | Tumors: (3) Normal tissues: (58) |
| $\sigma_f$ | Solvent-drag reflection coefficient | 0.83 | (16) |
| $\pi_v$ | Osmotic colloidal pressure in blood vessels | 15.4 mmHg | (65) |
| $\pi_i$ | Osmotic colloidal pressure in the interstitium | Tumors: 17.3 mmHg; Normal tissues: 11.0 mmHg | Tumors: (58); Normal tissues: (65) |
| $\pi_L$ | Osmotic colloidal pressure in the lymphatic vessels | Normal tissue: 11.8 mmHg | (65) |
| $\tau$ | tortuosity | 1.41 | (20) |
| $\varphi$ | Available volume fraction | Doxil: 0.05; bacteria: 0.05; Free doxorubicin: 0.26; | (18) |

| | |
|---|---|
| | 5-fluorouracil: 0.26; |